\documentclass[aps,amsfonts,pra,twocolumn]{revtex4}
\usepackage[unicode=true]{hyperref}
\usepackage{amssymb}
\usepackage{mathtools}
\usepackage[pdftex]{xcolor,graphicx}

\usepackage{braket}
\newtheorem{Theorem}{Theorem}

\newtheorem{Lemma}{Lemma}

\newtheorem{Def}{Definition}

\newcommand{\E}{E}

\begin{document}
\title{Phase space simulation method for quantum computation with magic states on qubits}
\author{$\text{Robert Raussendorf}^{1,2}$, $\text{Juani Bermejo-Vega}^{3}$, $\text{Emily Tyhurst}^4$, $\text{Cihan Okay}^{1,2}$, and $\text{Michael Zurel}^{1}$\vspace{4mm}\\
	{\small{\em{1: Department of Physics \& Astronomy, University of British Columbia, Vancouver, BC V6T1Z1, Canada}}} \vspace{0.2mm}\\
	{\small{\em{2: Stewart Blusson Quantum Matter Institute, University of British Columbia, Vancouver, BC, Canada}}}\vspace{0.2mm} \\
	{\small{\em{3: Freie Universit{\"a}t Berlin, Berlin, Germany}}}\vspace{0.2mm} \\
	{\small{\em{4: Department of Physics, University of Toronto, Toronto, ON, Canada}}}\vspace{0.2mm} \\
}

\begin{abstract}
We propose a method for classical simulation of finite-dimensional quantum systems, based on sampling from a quasiprobability distribution, i.e., a generalized Wigner function. Our construction applies to all finite dimensions, with the most interesting case being that of qubits. For multiple qubits, we find that quantum computation by Clifford gates and Pauli measurements on magic states can be efficiently classically simulated if the quasiprobability distribution of the magic states is non-negative. This provides the so far missing qubit counterpart of the corresponding result [V. Veitch {\em{et al.}}, New J. Phys. \textbf{14}, 113011 (2012)] applying only to odd dimension.  Our approach is more general than previous ones based on mixtures of stabilizer states. Namely, all mixtures of stabilizer states can be efficiently simulated, but for any number of qubits there also exist efficiently simulable states outside the stabilizer polytope. Further, our simulation method extends to negative quasiprobability distributions, where it provides probability estimation. The simulation cost is then proportional to a robustness measure squared. For all quantum states, this robustness is smaller than or equal to robustness of magic.
\end{abstract}

\maketitle

\section{Introduction}

How to mark the classical-to-quantum boundary is a question that dates back almost to the beginning of quantum theory. Ehrenfest's theorem \cite{Ehrenfest} provides an early insight, and the Einstein-Podolsky-Rosen paradox \cite{EPR} and Schr{\"o}dinger's cat \cite{Scat} are two early puzzles. The advent of quantum computation \cite{Feyn}--\cite{Deutsch} added a computational angle: When does it become hard to simulate a quantum mechanical computing device on a classical computer? Which quantum mechanical resource do quantum computers harness to generate a computational speedup?

One instructive computational model is quantum computation with magic states (QCM) \cite{BK}. In QCM, both ``traditional'' indicators of quantumness (developed in the fields of quantum optics and foundations of quantum mechanics) and a computational indicator can be applied. From quantum optics and foundations, the indicators are the negativity of a Wigner function \cite{Wig}--\cite{Buz}, and the breakdown of non-contextual hidden variable models \cite{Bell}--\cite{Merm}. Computer science is concerned with the breakdown of efficient classical simulation. 

In the particular setting of QCM, an important distinction arises between the cases of even and odd local Hilbert space dimension $d$. If $d$ is odd, then all three of the above indicators  for the classical-to-quantum boundary align \cite{NegWi}--\cite{Howard}. This is a very satisfying situation: the physicist, the philosopher and the computer scientist can have compatible notions of what is ``quantum''. 

In even local dimension, the situation differs starkly. Non-contextual hidden variable models for QCM are not viable regardless of computational power \cite{Howard}, which voids the foundational indicator, and furthermore obstructs the view of contextuality as a computational resource. Also, the multi-qubit Wigner functions constructed to date do not support efficient classical simulation of QCM by sampling over phase space. Thus, the physics and computer-science based criteria for classicality differ, which is an unsatisfactory state of affairs compared to odd $d$. The purpose of this paper is to align the perspectives of the physicist and the computer scientist on the classical-to-quantum transition in QCM on qubits.\medskip

To prepare for the subsequent discussion, we provide a short summary of QCM, and the role of the Wigner function in it. Quantum computation with magic states operates with a restricted set of instructions, the Clifford gates. These are unitary operations defined by the property that they map all Pauli operators onto Pauli operators under conjugation. Clifford gates are not universal, and, in fact, can be efficiently classically simulated \cite{Goma}. This operational restriction is compensated for by invoking the ``magic'' states, which are special quantum states that cannot be created by Clifford gates and Pauli measurements. Suitable magic states restore quantum computational universality; and in fact QCM is a leading paradigm for fault-tolerant universal quantum computation. In sum, computational power is transferred from the quantum gates to the magic states, and one is thus led to ask: Which quantum properties give the magic states their computational power?

One such property is, for odd $d$ at least, negativity in the Wigner function. A quantum speedup can arise only if the Wigner function of the magic states assumes negative values. If, to the contrary, the Wigner function is positive, then the whole quantum computation can be efficiently classically simulated \cite{NegWi},\cite{Mari}. Further, a positive Wigner function is, for $n\geq 2$ quantum systems, equivalent to the existence of a non-contextual hidden variable model \cite{Howard}, \cite{Delf2}.  Both Wigner function negativity and contextuality of the magic states are therefore necessary quantum computational resources.

As we noted, this picture only applies if the local Hilbert space dimension is odd. This excludes the full multi-qubit case, which arguably is the most important. Approaches to the qubit case have been made, e.g. through the rebit scenario \cite{ReWi} and multi-qubit settings with operational restrictions \cite{QuWi}, \cite{Bermejo}, or by invoking a Wigner function over Grassmann variables \cite{Love}, or multiple Wigner functions at once \cite{Galvao}. Common to these approaches is that, unlike for odd $d$ \cite{NegWi}, they do not efficiently simulate the evolution under general Clifford gates and Pauli measurements  by sampling, a.k.a. weak simulation \cite{VdN1}--\cite{BT}. 

An alternative approach to weak simulation is by defining a quasiprobability function over stabilizer states \cite{BK}, \cite{RoM}, \cite{Pashayan}, bypassing Wigner functions. It has the advantage of efficiently simulating all Clifford circuits on positively represented states. For multi-qubit systems, it has so far been unknown how the stabilizer method relates to Wigner functions, but we clarify the relation here.\medskip

In this paper, we provide the thus far missing phase space picture for QCM on multi-qubit systems. Central to our discussion is a new quasi-probability function defined for all local Hilbert space dimensions $d$ and all numbers of subsystems $n$. When applied to odd $d$, it reproduces the known finite-dimensional adaptation \cite{Gross}--\cite{Woott} of the original Wigner function \cite{Wig}; but for even $d$, in particular $d=2$, it is different. Then, this quasiprobability function requires a phase space of increased size, in accordance with \cite{KarWalBart}. Even in $d=2$, the positivity of this quasiprobability is preserved under all Pauli measurements. This property is crucial for the efficient classical simulation of QCM on positively represented states. Also, this simulation contains the efficient classical simulation \cite{BK} of stabilizer mixtures as a special case. We thus reproduce the essential features of the odd-dimensional scenario in $d=2$. 

Starting from the definition of the quasiprobability function $W$, we treat the following subjects: characterization of phase space for $d=2$, preservation of positivity of $W$ under Pauli measurements, covariance of $W$ under all Clifford unitaries, efficient classical simulation of QCM for $W\geq 0$, relation to the qubit stabilizer formalism, hardness of classical simulation for $W<0$, and a monotone under the free operations. 

In summary, we arrive at a description that resembles the corresponding scenario in odd local dimension. Namely, negativity in the quasiprobability distribution $W$ for the initial magic state is a necessary precondition for quantum speedup. However, one difference between even and odd $d$ remains. In odd $d$, every positive Wigner function is also a non-contextual hidden variable model. This is not so for even $d$, due to the phenomenon of state-independent contextuality among Pauli observables. 

\section{Results and outline}

\subsection{Summary of results}

This paper addresses the full $n$-qubit case of quantum computation with magic states, from the perspectives of the classical-to-quantum transition and quantum computational resources. For the case of local dimension  $d=2$ we closely reproduce the relations between Wigner function and efficient classical simulation existing in odd $d$. Central to our discussion is a novel quasiprobability function $W$ defined for all local Hilbert space dimensions $d$. It has the following general properties:

(i) For all $n$ and $d$, $W$ is Clifford-covariant and positivity-preserving under Pauli measurements.\smallskip

(ii) If the local Hilbert space dimension $d$ is even, $W_\rho$ is non-unique for any given quantum state $\rho$. The set of phase point operators corresponding to $W$ is over-complete.\smallskip

(iii) If $d$ is odd and $n\geq 2$, then $W$ reduces to the standard Wigner function \cite{Gross}, \cite{Gross2} for odd finite dimension. 
\smallskip

(iv) For all $n$ and $d$, the stabilizer formalism is contained as a special case. 
All stabilizer states can be positively represented by $W$, and efficiently updated under Clifford operations.\smallskip

(v) The present description goes beyond the stabilizer formalism. In particular, for $d=2$, for every number $n$ of qubits there exist non-mixtures of stabilizer states which are positively represented by $W$. Furthermore, for any quantum state $\rho$, the 1-norm of the optimal $W_\rho$ is smaller than or equal to the robustness of magic $\mathfrak{R}_S(\rho)$.  (Both robustness measures are instances of sum negativity \cite{Pashayan}.)
\smallskip

The following properties of $W$ for special values of $n$ (and $d=2$) are also worth noting. (a) The Eight-state model \cite{EightState} is a special case of $W$, namely for $n=1$. (b)
For Mermin's square \cite{Merm}, the present simulation algorithm saturates the lower bound \cite{MemCo} on the memory cost of classical simulation. (c) Up to two copies of magic $T$ and $H$ states are positively represented by $W$.\medskip

We establish the following main results: (I) The set of states positively represented by $W$ is closed under Pauli measurement (Theorem~\ref{CPM} in Section~\ref{QUP}). (II) If a quantum state $\rho$ has a non-negative function $W_\rho$, and $W_\rho$ can be efficiently sampled, then, for every Clifford circuit applied to $\rho$, the corresponding measurement statistics can be efficiently sampled (Theorem~\ref{T1} in Section~\ref{CEsim}). In this sense, $W\geq 0$ leads to efficient classical simulation of the corresponding quantum computation. (III) For $d=2, n\geq 2$, the $n$-system phase space has a more complicated structure than in the case of odd $d$, reflecting the fact that the phase point operators are dependent. The points in generalized multi-qubit phase space are classified (Theorem~\ref{nms} in Section~\ref{Struct}). (IV) There exists a robustness measure $\mathfrak{R}$ which bounds the hardness of classical simulation of quantum computation with magic states, when $W_{\rho_\text{init}}<0$ for the initial state $\rho_\text{init}$. $\mathfrak{R}$ is less than or equal to the robustness of magic (Lemma~\ref{relR}), and a monotone under Clifford unitaries and Pauli measurements (Theorem~\ref{Mono} in Section~\ref{CEsimNeg}).

\subsection{Outline}

The remainder of this paper is organized as follows. In Section~\ref{sim} we define a quasiprobability function $W$. We show that it reduces to Gross' Wigner function \cite{Gross} whenever the local Hilbert space dimension $d$ is odd, but, more importantly, is different in even dimension. Specifically, $W$ represents all quantum states redundantly for even $d$, which enables Clifford covariance and positivity preservation under Pauli measurement. In Section~\ref{Struct} we analyze the structure of the phase space on which $W$ lives, for the case of multiple qubits. In particular, we classify the points of phase space. We also clarify the relation to the qubit stabilizer states and their mixtures.

In Sections~\ref{QUP} and \ref{CEsim} we turn to dynamics. In Section~\ref{QUP} we discuss the update of $W$ under Pauli measurement, and in Section~\ref{CEsim} the efficient classical simulation of QCM for positive $W$. 

In Section~\ref{CEsimNeg} we address the case of $W_\rho<0$. We discuss hardness of classical simulation, as well as the elements of a resource theory based on $W$. 

In Section~\ref{Disc} we discuss the extent to which the quasiprobility function $W$ satisfies the Stratonovich-Weyl criteria, and its relation to hidden variable models.
We conclude in Section~\ref{Concl}.

\section{The quasiprobability function}\label{sim}

In this section we introduce the generalized $n$-qudit phase space ${\cal{V}}$, for any local Hilbert space dimension $d$, and a quasi-probability distribution $W:{\cal{V}} \longrightarrow \mathbb{R}$ living on it. In Section \ref{PhaseSp} we define the phase point operators corresponding to $W$, and in Section \ref{Maxsets} identify a minimal set of them. Section \ref{coho} reveals the cohomological underpinning of our construction, which links the present subject to parity proofs of quantum contextuality \cite{Coho} and contextuality in measurement-based quantum computation~\cite{CohoMBQC}. 

\subsection{Generalized phase space}\label{PhaseSp}

We choose a phase convention for the Pauli operators,
\begin{equation}\label{Pauli}
	T_a = e^{i\phi(a)}X(a_X)Z(a_Z),\;\; \forall a=(a_X,a_Z) \in \E:=\mathbb{Z}_d^{2n}.	
\end{equation}
Therein, the function $\phi:\E \longrightarrow \mathbb{R}$ has to satisfy the constraint that  $(T_a)^d=I$, for all $a\in \E$. As a consequence of this condition, all eigenvalues of the operators $T_a$ are of the form $\omega^k$, $k\in \mathbb{N}$, with $\omega:=\exp(2\pi i/d)$.

We now proceed to the definition of the phase point operators. We consider a subset $\Omega$ of $\E$, and a function $\gamma: \Omega \longrightarrow \mathbb{Z}_d$, both subject to additional constraints that will be specified in Definitions~\ref{Def_Cl}--\ref{PhaSpa} below. The pair $(\Omega,\gamma)$ specifies a corresponding phase point operator $A_\Omega^\gamma$,
\begin{equation}\label{PPO2}
A_\Omega^\gamma := \frac{1}{d^n}\sum_{b \in \Omega}\omega^{\gamma(b)}T_b,
\end{equation}
with the constraint that
\begin{equation}\label{NormCond}
	\omega^{\gamma(0)}T_0 = I.
\end{equation}
When comparing Eq.~(\ref{PPO2}) to the phase point operators of the previously discussed qudit \cite{NegWi}, rebit \cite{ReWi} and restricted qubit \cite{QuWi} cases, we note that the overall structure remains the same. In this case, the sets $\Omega$ are an additional varying parameter, and the phase space thereby becomes larger.

Based on the phase point operators $A_\Omega^\gamma$ of Eq.~(\ref{PPO2}), we introduce the counterpart to the Wigner function that applies to our setting. The generalized phase space ${\cal{V}}$ consists of all admissible pairs $(\Omega,\gamma)$, to be specified below. Any $n$-system quantum state $\rho$ can be expanded in terms of a function $W_\rho: {\cal{V}} \longrightarrow \mathbb{R}$,
\begin{equation}\label{Wigner}
\rho = \sum_{(\Omega,\gamma) \in {\cal{V}}}W_\rho(\Omega,\gamma) A_\Omega^\gamma.
\end{equation}
The reason for imposing Eq.~(\ref{NormCond}) is that it implies $\text{Tr}A_\Omega^\gamma = 1$, for all $(\Omega,\gamma)\in {\cal{V}}$. Hence, $W$ defined in Eq.~(\ref{Wigner}) is a quasiprobability distribution. As we see shortly, it generalizes the Wigner function \cite{Gross} for odd-dimensional qudits to qubits. 

We note that when $d$ is even, the quasiprobability distribution $W_\rho$ is non-unique because the set of phase point operators of Eq.~(\ref{PPO2}) is overcomplete. 
\begin{Def}
An $n$-qudit quantum state $\rho$ is {\em{positively representable}} if it can be expanded in the form of Eq.~(\ref{Wigner}), with $W_\rho(\Omega,\gamma)\geq 0$, for all $(\Omega,\gamma) \in {\cal{V}}$.
\end{Def}

The efficient classical simulation algorithm described in Section~\ref{CEsim} applies to positively representable quantum states $\rho$. The non-uniqueness of $W_\rho$ allows for more positively representable states than prior quasiprobability representations.\medskip

We now turn to the properties of admissible sets $\Omega$ and functions $\gamma$ that define points in the phase space ${\cal{V}}$. To begin, we define a function $\beta$ which encodes how translation operators on phase space compose, 
\begin{align}\label{3T}
T_aT_b = \omega^{\beta(a,b)}T_{a+ b},\,\forall a,b \in \E\text{ : } [T_a,T_b]=0.
\end{align}
We further define the symplectic product
\begin{align}
[a,b]:= a_Xb_Z-a_Zb_X\mod d,
\end{align}
and hence $[a,b]=0\;\Longleftrightarrow\; [T_a,T_b]=0$. 

The function $\beta$ satisfies the relation
\begin{equation}\label{beta_constr}
\beta(a,b)+\beta(a+b,c)-\beta(b,c)-\beta(a,b+c)=0\!\! \mod d,
\end{equation}
for $ a,b,c\in \E$. We state this relation for later reference. It is a consequence of the associativity of operator multiplication. Consider the operator product $T_aT_bT_c=T_a(T_bT_c)=(T_aT_b)T_c$, and expand $T_a(T_bT_c) = \omega^{\beta(a,b+c)+\beta(b,c)}T_{a+b+c}$, $(T_aT_b)T_c=\omega^{\beta(a,b)+\beta(a+b,c)}T_{a+b+c}$. Comparing the two equivalent expressions yields Eq.~(\ref{beta_constr}).

Then, it follows straightforwardly from the definition Eq.~(\ref{3T}) of $\beta$ that
\begin{equation}\label{betaSim}
\beta(a,b) = \beta(b,a),\;\;\forall a,b \;\text{with}\; [a,b]=0. 
\end{equation}
We constrain $\Omega$ by the following definitions:
\begin{Def}\label{Def_Cl}
	A set $\Omega \subset \E$ is {\em{closed under inference}} if it holds that
	\begin{equation}\label{cui}
	a,b \in \Omega\, \wedge\, [a,b]=0 \Longrightarrow a+ b \in \Omega.
	\end{equation}
\end{Def}
The motivation for this definition is that if $T_a$ and $T_b$ can be simultaneously measured, then the value of $T_{a+ b}$ can be inferred from the measurement outcomes, through relation (\ref{3T}). A consequence of the closedness under inference is that $0 \in \Omega$ for all closed sets $\Omega$.

\begin{Def}\label{Def_NC}
	A set $\Omega \subset \E$ is {\em{non-contextual}} if there exists a value assignment $\gamma: \Omega\longrightarrow \mathbb{Z}_d$ that satisfies the condition
	\begin{equation}\label{dgb}
	\gamma(a)+\gamma(b)-\gamma(a+ b) =\beta(a,b), 
	\end{equation}
	for all $a,b \in \Omega$, and $[a,b]=0$. 
\end{Def}
To motivate the nomenclature, if the set $\Omega \subset \E$ is non-contextual per the above definition, then it does not admit a parity-based contextuality proof \cite{Coho}. Namely, Eq.~(\ref{dgb}) represents the constraints on non-contextual value assignments $\gamma$ that result from the operator constraints Eq.~(\ref{3T}). If these constraints can be satisfied, then there is no parity-based contextuality proof.

\begin{Def}\label{PhaSpa} The generalized phase space ${\cal{V}}$ consists of all pairs $(\Omega,\gamma)$ such that
	(i) $\Omega$ is closed under inference, (ii) $\Omega$ is non-contextual, (iii) $\gamma: \Omega \longrightarrow \mathbb{Z}_d$ satisfies the relation Eq.~(\ref{dgb}), and (iv) Eq.~(\ref{NormCond}) holds.
\end{Def} 
Thus, for the generalized phase space ${\cal{V}}$, the only sets $\Omega$ that matter are simultaneously closed and non-contextual. For short, we call such sets ``cnc''. 
	
\subsection{Maximal sets $\Omega$}\label{Maxsets}
	
The cnc sets $\Omega$ partially specify the points in phase space, and it is thus desirable to eliminate possible redundancies among them. It turns out that only the ``maximal'' sets $\Omega$ need to be considered for ${\cal{V}}$.

\begin{Def}
	A cnc set $\Omega \subset \E$ is maximal if there is no cnc set $\tilde{\Omega}\subset \E$ such that $\Omega \subsetneq \tilde{\Omega}$.
\end{Def}
We denote by ${\cal{V}}_M$ the subset of ${\cal{V}}$ constructed only from the maximal cnc sets $\Omega$. Then, any quantum state $\rho$ has expansions like Eq.~(\ref{Wigner}), but with ${\cal{V}}$ replaced by ${\cal{V}}_M$. If one of those expansions is non-negative, then we say that $\rho$ is positively representable w.r.t. ${\cal{V}}_M$.
	
	\begin{Lemma}\label{MaxOm}
		For any $n$ and $d$, a quantum state $\rho$ is positively representable w.r.t. ${\cal{V}}$ if and only if it is positively representable w.r.t. ${\cal{V}}_M$. 
	\end{Lemma}
	From the perspective of positive representability, we may therefore shrink ${\cal{V}}$ to ${\cal{V}}_M$ without loss. We make use of this property when discussing the case of odd $d$ in Section~\ref{Odd_d} right below, and in the classification of cnc sets $\Omega$ for the multi-qubit case in Section~\ref{Gen}. The proof of Lemma~\ref{MaxOm} is given in Appendix~\ref{MaxOmProof}.

\subsection{The cohomological viewpoint \label{coho}} 

The above Definitions~\ref{Def_NC} and \ref{PhaSpa} have a cohomological underpinning, which connects the subject of the present paper to the topological treatment of parity-based contextuality proofs \cite{Coho}, and of contextuality in measurement-based quantum computation \cite{CohoMBQC}.

The cohomological picture arises as follows. The partial value assignments $\gamma$ and the function $\beta$ are cochains in a chain complex, with Eqs.~(\ref{beta_constr}) and (\ref{dgb}) constraining them.  Eq.~(\ref{beta_constr}) says that $\beta$ is a special cochain, namely a cocycle. Now, the basic reason for why the case of even $d$ is so much more involved than the case of odd $d$ is that, for even $d$, the cocycle $\beta$ is non-trivial whereas for odd $d$ it is trivial \cite{Coho}.

Eqs.~(\ref{beta_constr}) and (\ref{dgb}) are frequently used in this paper, for example in the update rules of the phase point operators under Pauli measurements (proof of Lemma~\ref{Q_up}), the closedness of the generalized phase space ${\cal{V}}$ under update by Pauli measurement (proof of Lemma~\ref{Perpet}), and covariance of the quasiprobability function $W$ under Clifford unitaries (proof of Lemma~\ref{Covar}). These are central properties for the phase-space description of quantum computation with magic states. and they are all matters of cohomology. \smallskip

The cohomological formulation is based on a chain complex ${\cal{C}}_n$ constructed from the $n$-qubit Pauli operators $T_a$. The operator labels $a$ define the edges of this complex; the faces of ${\cal{C}}_n$ correspond to commuting pairs $(a,b)$ and volumes $(a,b,c)$ to commuting triples. For details, the interested reader is referred to \cite{Coho}. Here, we only state two basic topological properties of the present scenario. 

As already noted, the cochain $\beta$ defined in Eq.~(\ref{3T})  is in fact a 2-cocycle, with the cocycle condition $d\beta = 0$ enforced by Eq.~(\ref{beta_constr}). For any given volume $v=(a,b,c)$, the coboundary $d\beta$ evaluates on $v$ to
\begin{equation}\label{Def_deb}
d\beta(a,b,c):=\beta(a,b)+\beta(a+b,c)-\beta(b,c)-\beta(a,b+c).
\end{equation}
Thus, Eq.~(\ref{beta_constr}) says that $d\beta(v) =0$, for all volumes $v$.

Eq.~(\ref{dgb}) in Definition~\ref{Def_NC} also has a cohomological interpretation, namely  $d\gamma =  \beta|_{\Omega\times \Omega}$, with
\begin{equation}\label{Def_deg}
d\gamma(a,b):=\gamma(a)+\gamma(b)-\gamma(a+b),
\end{equation}
for any face $(a,b)$ spanned by commuting edges $a,b\in \E$.

Subsequently, we use evaluations of $d\beta$ and $d\gamma$, defined in Eqs.~(\ref{Def_deb}) and (\ref{Def_deg}), as a short-hand to express Eqs.~(\ref{beta_constr}) and (\ref{dgb}).  As outlined above, it is conceptually helpful to remember that $d\beta$ and $d\gamma$ denote coboundaries, but it is not required for the technical results presented in this paper.

\section{Properties of the phase space ${\cal{V}}$}\label{Struct}

In this section, we look at the structure of the phase space ${\cal{V}}$ more closely, and make connections to previous phase space formulations. Namely, in Section~\ref{Odd_d} we address the relationship of this phase space to the usual qudit phase space, and in Section~\ref{QR} we make clear the connections to the previously addressed rebit case. Further, in Section~\ref{Gen}, we classify the cnc sets $\Omega$, and for every $\Omega$ describe the sets $\Gamma(\Omega)$  of value assignments $\gamma$. In Section~\ref{Exa} we clarify the relation to the stabilizer formalism. 

\subsection{Qudits of odd dimension}\label{Odd_d}

This is the only place in the present paper where we consider the case of odd $d$. The purpose of this section is to show that if $d$ is odd then for $n\geq 2$ qudits the generalized phase space ${\cal{V}}$ reduces to the standard phase space $V=\mathbb{Z}_d^{2n}$. There, the quasiprobability function $W$ becomes the standard Wigner function \cite{Gross} for odd finite-dimensional systems. Hence, the present quasiprobability function $W$ is a generalization of the finite odd dimensional Wigner function \cite{Gross}, which in turn is a descendant of the original Wigner function \cite{Wig}.
\medskip

If $d$ is odd then the whole set $\E$ is cnc. First, $\E$ is closed under inference by definition. And second, it is known that in odd dimension Pauli observables have non-contextual deterministic value assignments \cite{Howard2}, \cite{QuWi}. These yield the functions $\gamma$, satisfying the condition Eq.~(\ref{dgb}). $\E$ is thus non-contextual.

$\E$ is furthermore the single maximal set, and, with Lemma~\ref{MaxOm}, the only cnc set that needs to be considered for the phase space. Hence, the phase point operators are
$$
A_\E^\gamma =\frac{1}{d^n}\sum_{a\in E} \omega^{\gamma(a)} T_a,
$$
with the functions $\gamma$ satisfying Eqs.~(\ref{NormCond}) and (\ref{dgb}). The former condition ensures that the identity operator appears with weight $1/d^n$ in the expansion (real and positive). If $n\geq 2$, the latter condition has $d^{2n}$ solutions for the functions $\gamma$ if $d$ \cite{Delf2}. For a suitable choice of $\phi$ in Eq.~(\ref{Pauli}), it holds that $\beta \equiv 0$ (odd $d$ only). The solutions for $\gamma$ then form a vector space
$$
V = \mathbb{Z}_d^{2n}\;\;\;\; \text{(for odd $d$)}.
$$
We note that the case of a single qudit, $n=1$, is an exception to the above behaviour. In this case, the set $\mathcal{V}$ has greater cardinality than $\mathbb{Z}_d^2$ \cite{Delf2}. 

\subsection{Qubits and rebits}\label{QR}

The remainder of this paper is about local Hilbert space dimension $d=2$. This means mostly qubits, but we will occasionally also consider systems of rebits. The reason is that the major complication of the $d=2$ case stems from Mermin's square and star \cite{Merm}---two strikingly simple contextuality proofs. Those settings embed most efficiently in rebits rather than qubits, which warrants the inclusion of rebits here.

We remark that the present discussion of rebits is almost identical to the discussion of qubits, but very different from the earlier discussion of rebits in \cite{ReWi}. In the latter, the physically measurable observables were restricted from real Pauli operators to CSS-type Pauli operators, and the real Clifford unitaries to CSS-ness preserving Clifford unitaries. No such restrictions are imposed here. If the restriction to CSS-ness preserving operations is imposed, then Mermin's square and star, along with all other state-independent contextuality proofs based on Pauli observables, are effectively excised \cite{ReWi}. Here, we retain those contextuality proofs, and consequently have to adjust to their presence. Notably, these contextuality proofs constrain quasiprobability distributions that preserve positivity under Pauli measurement.

We start the exploration of the $d=2$ case with two examples that illustrate the concept of generalized phases space ${\cal{V}}$. The second example also illustrates the differences between contextuality, negative quasiprobability and quantum computational power for two-level systems.\medskip

{\em{Example 1: Eight-state model.}} It is known that {\em{every}} one-qubit quantum state can be positively represented by the so-called Eight-state-model \cite{EightState}, which consists of two standard 1-qubit Wigner functions tagged together. The Eight-state-model is an instance of the state expansion Eq.~(\ref{Wigner}), namely for $d=2$, $n=1$, and it contains only one set $\Omega$,
$$
\Omega_0=\{0,x,y,z\},
$$
with $T_0=I$, $T_x=X$, $T_y=Y$ and $T_z=Z$. It is easily checked that $\Omega_0$ is non-contextual and closed under inference (no inference possible). The value assignments $\gamma$ are constrained by Eq.~(\ref{NormCond}), hence $\gamma(0)=0$, and no constraints arise from Eq.~(\ref{beta_constr}) due to the lack of non-trivial commuting elements in $\Omega_0$. Thus, $\gamma(x)$, $\gamma(y)$ and $\gamma(z)$ can be freely chosen. There are eight resulting functions, and they define the eight states of the model. 

All one-qubit quantum states can be positively represented by this model, which is strictly more than all mixtures of one-qubit stabilizer states.\medskip

\begin{figure}
	\begin{center}
		\begin{tabular}{lll}
			(a) & (b) & (c)\vspace{2mm}\\
			\includegraphics[width=2.2cm]{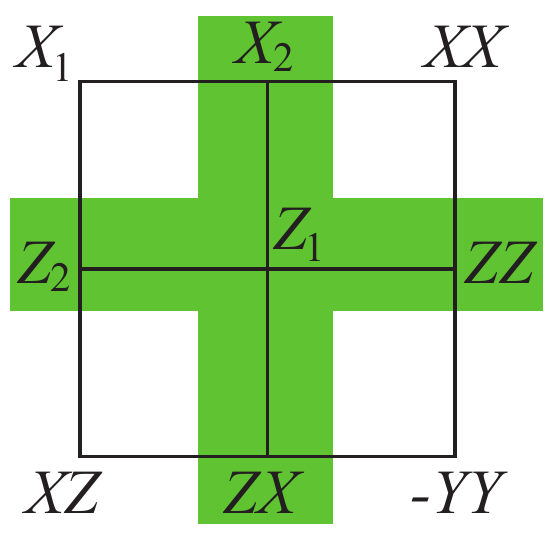} \hspace*{3mm}& \includegraphics[width=2.2cm]{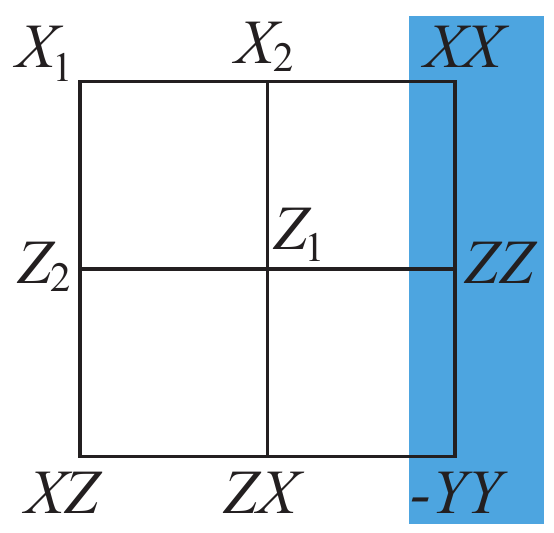}  \hspace*{3mm} & 
			\includegraphics[width=2.2cm]{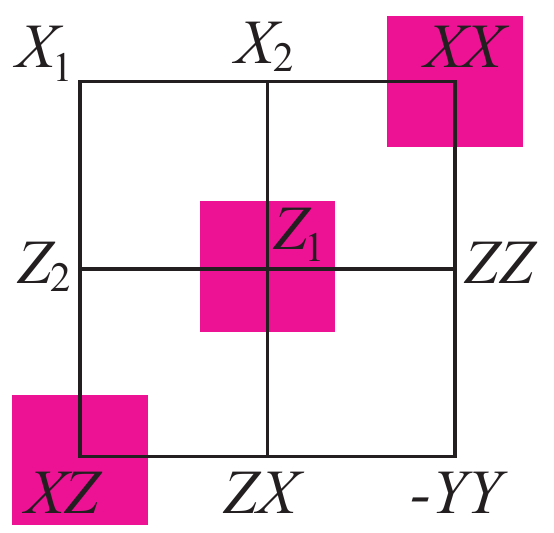}
		\end{tabular}
		\caption{\label{MermOm} Three types of cnc sets $\Omega$ for Mermin's square. (a) union of two isotropic subspaces intersecting in one element, (b) isotropic subspace, (c) triple of anti-commuting elements.}
	\end{center}
\end{figure}

{\em{Example 2: Mermin's square.}} Mermin's square is at the very root of the complications that arise for Wigner functions in even dimension. In particular, no $n$-qubit Wigner function for which the corresponding phase point operators form an operator basis can preserve positivity under all Pauli measurements \cite{QuWi}. 

All observables appearing in Mermin's star are real, and can thus be embedded in two rebits. Our formalism is easily adaptable to this slightly simpler scenario. Fig.~\ref{MermOm} shows three distinct types of cnc sets $\Omega$. Type (a) is the union of two non-trivially intersecting isotropic subspaces (9 sets), type (b) is isotropic subspaces (6 sets), and type (c) is triples of anti-commuting elements, i.e., one from each row and column of the square (6 sets). For each  cnc set $\Omega$ of type (a), (b) and (c) of Fig.~\ref{MermOm}, the constraint Eq.~(\ref{dgb}) allows for $2^3$, $2^2$, $2^3$ functions $\gamma$, respectively. The number of phase space points of each type therefore is 72, 24, 48.

We make the following numerical observations about the two-rebit case: (i) Random sampling suggests that all 2-rebit states are positively representable;  see Table~\ref{Vol}. (ii) In Fig.~\ref{fig:crossSection} the region of positively representable density matrices of the form 
\begin{equation}\label{Sect}
\begin{array}{rcl}
\rho(x,y)&=&\displaystyle{\frac{1}{4}I_{12}+x(X_1X_2+Z_1Z_2-Y_1Y_2)}\vspace{1mm}\\
&&\displaystyle{+y(Z_1+Z_2),}
\end{array}
\end{equation}
for $x,y \in \mathbb{R}$, is displayed for three different methods; namely the stabilizer method \cite{RoM}, the hyper-octrahedral method \cite{Rall}, and the present phase space method. We find that all quantum states in the plane spanned by the parameters $x$, $y$ are positively represented by the present phase space method, and this is not the case for the stabilizer and hyper-octahedral methods. \smallskip

{{\em{Example 3: 2 qubits.}} Numerical analysis shows that two copies of the state 
	\begin{equation}\label{Tphi}
	|H(\phi)\rangle:=(|0\rangle + e^{-i\phi}|1\rangle)/\sqrt{2}
	\end{equation} 
	can be positively represented, for all angles $\phi$.

	\begin{figure}
		\begin{center}
			\begin{tabular}{l}(a) \\ \includegraphics[width=0.25\textwidth]{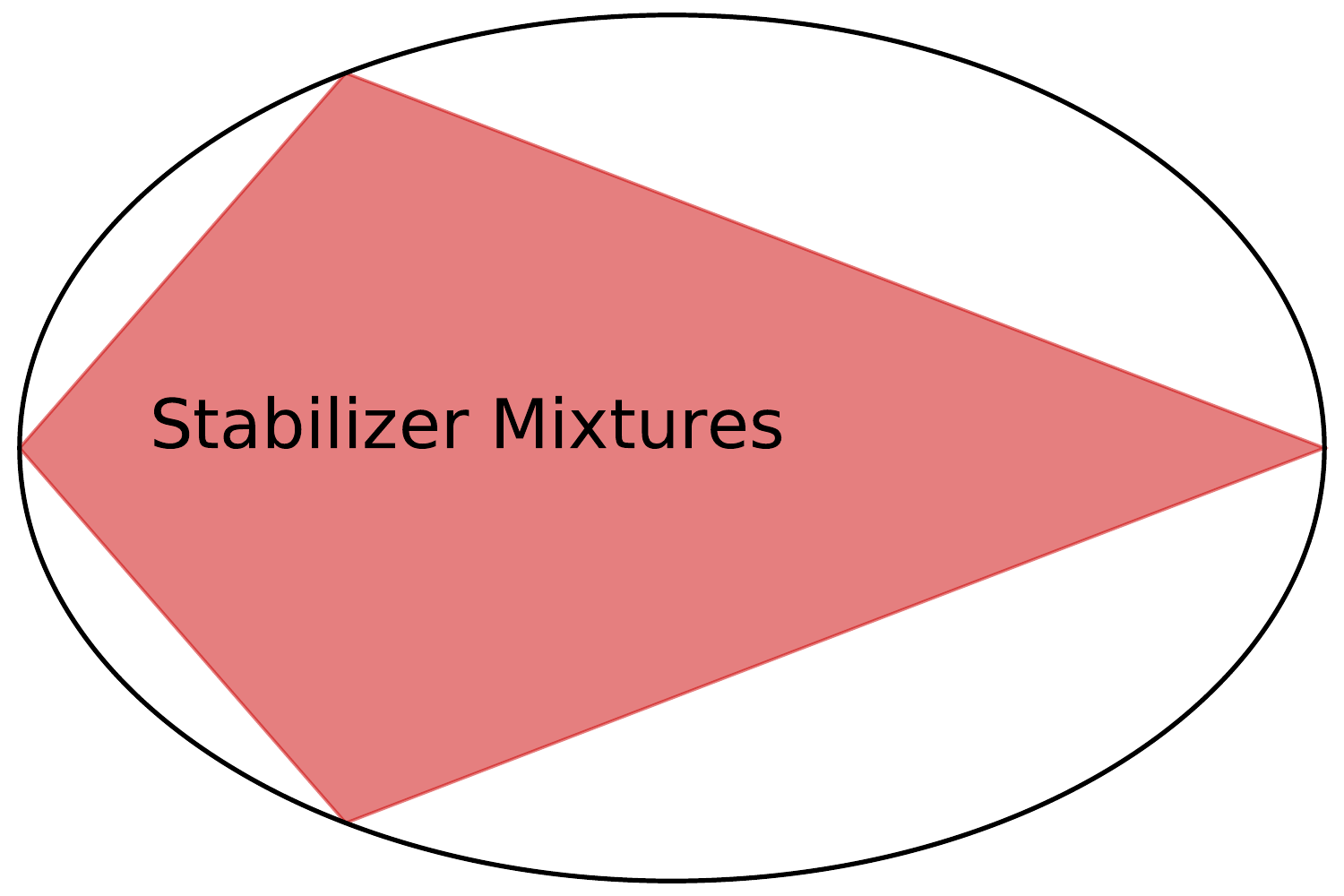}\vspace{3mm}\\
				
				(b)\\ \includegraphics[width=0.25\textwidth]{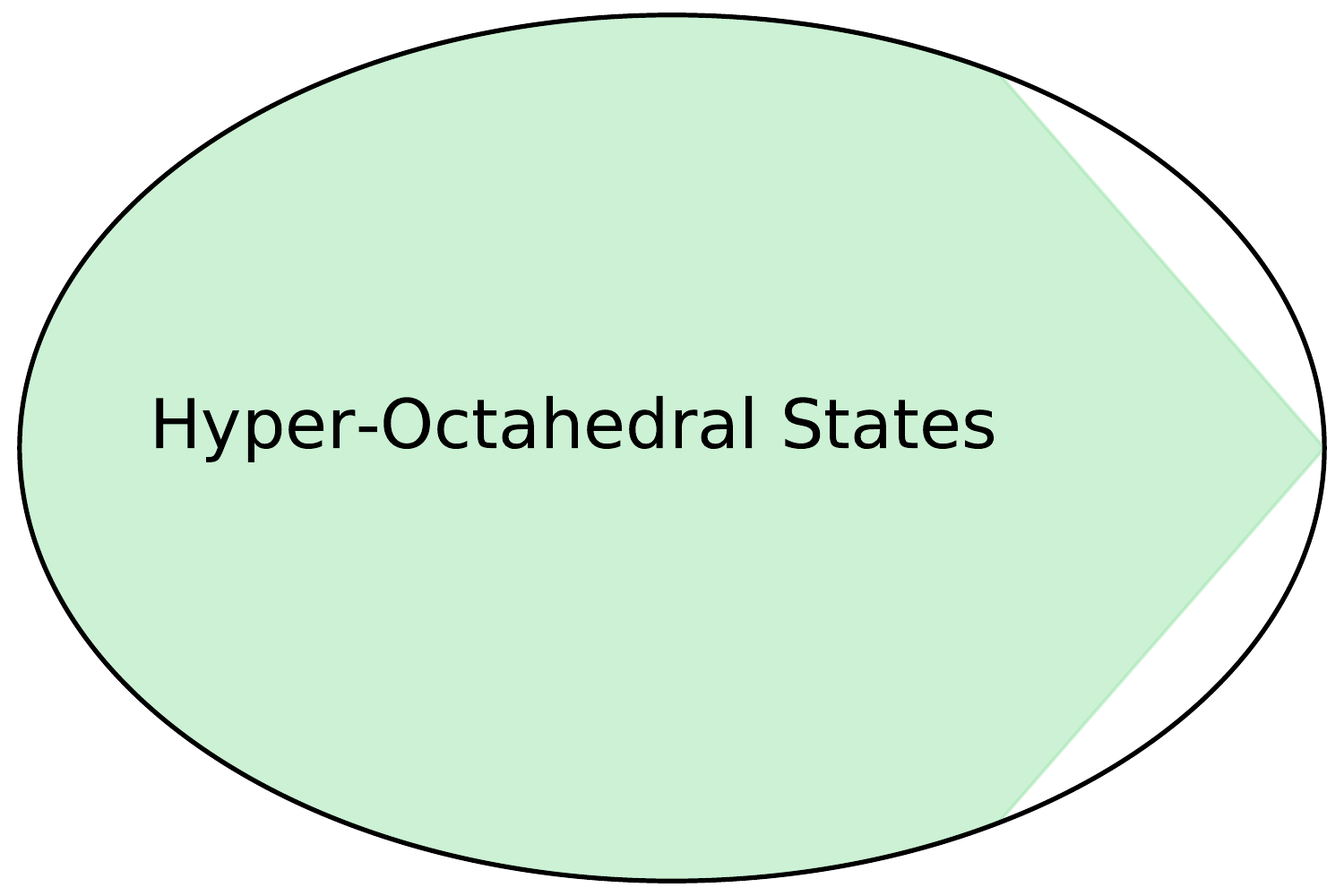}\vspace{3mm}\\
				
				(c)\\ \includegraphics[width=0.25\textwidth]{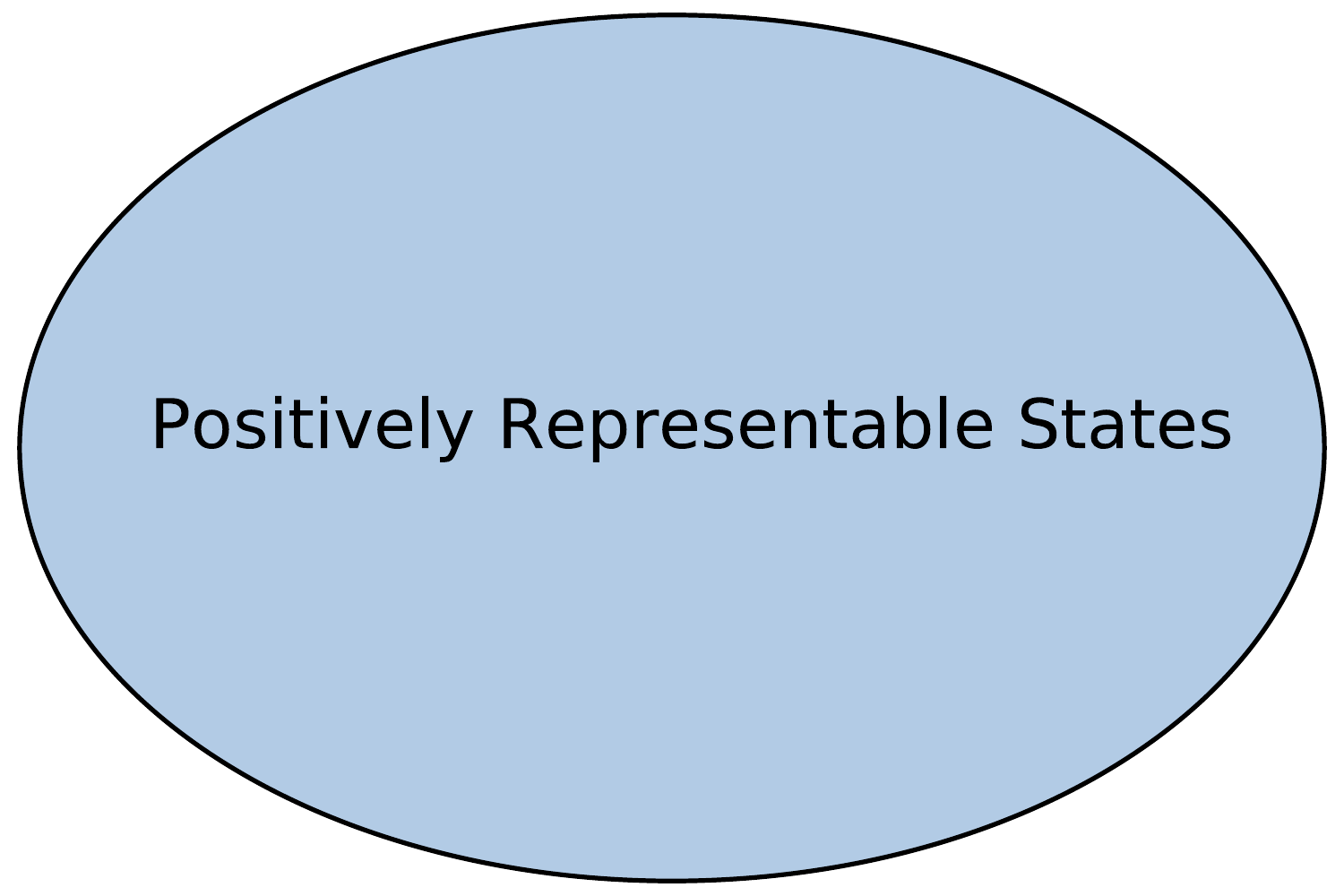}
			\end{tabular}
			\caption{\label{fig:crossSection}Two-dimensional cross section of the two-qubit state space, as parameterized in Eq.~(\ref{Sect}). The shaded regions indicate the positively representable states by various methods; (a) mixtures of stabilizer states, (b) hyper-octahedral states~\cite{Rall}, and (c) states positively represented by the present phase space method.}
		\end{center}
	\end{figure}

\subsection{Classification of multi-qubit phase space points}\label{Gen}

Denote by $\Gamma(\Omega)$ the set of functions $\gamma:\Omega \longrightarrow \mathbb{Z}_2$ that satisfy the constraints Eqs.~(\ref{NormCond}) and (\ref{dgb}). Then, the following statement holds.

\begin{Lemma}\label{coset1}
For all sets $\Omega$ of Def.~\ref{PhaSpa}, $\Gamma(\Omega)$ is the coset of a vector space $U(\Omega)$.  
\end{Lemma}

{\em{Proof of Lemma~\ref{coset1}.}} Write $\gamma = \gamma_0 + \eta$, where $\gamma_0\in \Gamma(\Omega)$ is some reference function. Then, the only condition on the functions   $\eta\in U(\Omega)$ is $d\eta =0$. Thus, if $\eta,\eta'\in U(\Omega)$ then $c\eta+c'\eta' \in U(\Omega)$, for all $c,c' \in \mathbb{Z}_2$. $\Box$\medskip

Lemma~\ref{coset1} reproduces a familiar feature. In infinite and finite odd dimension, the whole phase space is an orbit under the vector space of translations. There is an origin 0 of phase space, and all other phase space points are obtained from it by translation. In our present case of $d=2$, the phase space ${\cal{V}}$ splinters into many fragments, each of which corresponds to a vector space $U$ attached to a cnc set $\Omega$. 

At this point, one question about the structure of ${\cal{V}}$ remains: Can the cnc sets $\Omega$ be classified? It is resolved by Lemma~\ref{L4} and Theorem~\ref{nms} below.

\begin{Lemma}\label{L4}
For $n$ qubits, consider an isotropic subspace $\tilde{I} \subset E$ of dimension $n-m$, with $m \leq n$, and $\xi \leq 2m+1$ elements $a_k \in E$ that pairwise anti-commute but all commute with $\tilde{I}$. Denote $I_k := \langle a_k,\tilde{I}\rangle$ for $k=1,..,\xi$. For any number $n$ of qubits, the sets 
\begin{equation}\label{manyI}
\Omega = \bigcup_{k=1}^{\xi} I_k
\end{equation}
are non-contextual and closed under inference.
\end{Lemma}
\medskip

\begin{figure}
\begin{center}
\begin{tabular}{cccc}
	\includegraphics[width=2.1cm]{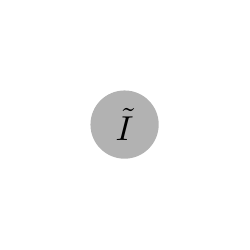} &	
	\includegraphics[width=2.1cm]{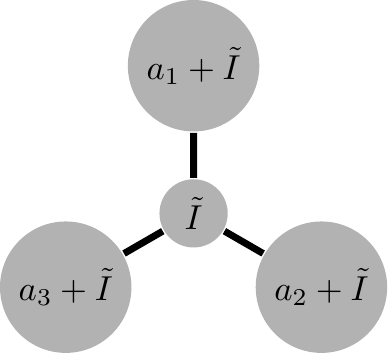} &\hspace*{4mm}&
	\includegraphics[width=2.1cm]{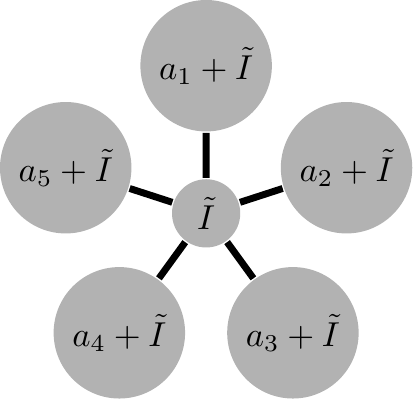}\\
		$m=0$ & $m=1$ && $m=2$
\end{tabular}
\caption{Commutativity graph representation for the cosets of Eq.~(\ref{manyI}) sets.  Elements pair-wise commute within each vertex and elements in adjacent vertices pair-wise commute.  Elements in non-adjacent vertices anti-commute.}
\end{center}	
\end{figure}

{\em{Proof of Lemma~\ref{L4}.}} {\em{Existence.}} The sets $\Omega$ of Eq.~(\ref{manyI}) exist for all $m$, $n$. To see this, 
consider the $m$-qubit Jordan-Wigner transforms of the Majorana fermion operators acting on qubits 1 to $m$,
\begin{equation}\label{Majorana}
\begin{array}{rcl}
C_{2j-1} &=& I_{1..j-1} X_jZ_{j+1}Z_{j+2}..Z_{m-1}Z_m,\\ 
C_{2j} &=& I_{1..j-1} Y_jZ_{j+1}Z_{j+2}..Z_{m-1}Z_m,
\end{array}
\end{equation}
for $j=1,..,m$, and, if $m>0$, the further observable
\begin{equation}\label{MajoB}
C_{2m+1} = Z_1Z_2... Z_{m-1}Z_m.
\end{equation}
Further, be $\tilde{I}$ the isotropic subspace corresponding to a stabilizer state supported on the $n-m$ qubits numbered $m+1, .. ,n$. Define $a_k$ via $C_k=T_{a_k}$ as in Eqs.~(\ref{Majorana}),(\ref{MajoB}), for all $k=1,..,2m+1$. These $a_k$ and $a\in\tilde{I}$ have the commutation relations required. \smallskip

{\em{Closedness.}} Consider a pair $c,d\in \Omega$ such that $[c,d]=0$. There are two cases. (i) $c,d\in I_k$, for some $k$. Then, $c+d \in I_k$, hence $c+d \in \Omega$.

(ii) $c\in I_k$ and $d \in I_l$, $k\neq l$. We may write $c=\nu\, x + g$, $d=\mu\, y+ g'$, for some $\nu,\mu \in \mathbb{Z}_2$ and $g,g'\in \tilde{I}$. The commutation relation $[c,d]=0$ then implies that $\nu\mu=0$, hence either $\nu=0$ or $\mu=0$. Wlog. assume that $\nu=0$. Then, $c\in \tilde{I}$, hence $c,d \in I_l$. Thus, $c+d \in I_l \subset \Omega$.

In both cases, $c,d\in \Omega$ and $[c,d]=0$ implies that $c+d\in \Omega$. Hence, $\Omega$ is closed under inference.\smallskip

{\em{Non-contextuality.}} There exists a function $\gamma|_{\tilde{I}}: \tilde{I} \longrightarrow \mathbb{Z}_2$ that satisfies Eq.~(\ref{dgb}) on $\tilde{I}$. We now extend this function to $\Omega$ as follows. The values $\gamma(a_k)$, for $k=1,..,\xi$ can be freely chosen, and for all $a\in \tilde{I}$ and all $k$, $\gamma(a_k+a):= \gamma(a_k)+\gamma(a)+\beta(a_k,a)$. This fully defines $\gamma:\Omega\longrightarrow \mathbb{Z}_2$. All commuting triples $c,d,c+d$ lie within one of the isotropic spaces $I_k$ forming $\Omega$, and $d\gamma(a,b)=\beta(a,b)$ thus holds.\smallskip 

This establishes that the sets $\Omega$ of Eq.~(\ref{manyI}) exist for the maximum value of $\xi$, $\xi=2m+1$. One may always choose $\xi$ smaller, which neither affects closedness nor non-contextuality. 
$\Box$\medskip

\begin{Theorem}\label{nms}
	All maximal cnc sets $\Omega$ are of the form Eq.~(\ref{manyI}), with $\xi=2m+1$ and $1\le m\le n$.
\end{Theorem}

{\em{Proof of Theorem~\ref{nms}.}}  Let $\Omega\subset E$ be closed under inference and non-contextual.  We can partition the elements of $\Omega$ into two subsets, $\Omega=\{q_1,\dots,q_\mu|g_1,\dots,g_\nu\}$, where $\tilde{I}=\{g_1,\dots,g_\nu\}$ are the elements of $\Omega$ which commute with the whole set.  $\tilde{I}$ is an isotropic subspace since if two elements, $a$ and $b$, commute with $\Omega$, then clearly their sum, $a+b$, also commutes with $\Omega$, and $\tilde{I}$ is isotropic by definition.
	
	If all elements of $\Omega$ pair-wise commute then $\Omega=\tilde{I}$ is an isotropic subspace.  Isotropic subspaces are not maximal cnc sets because they are always contained in Eq.~(\ref{manyI}) sets with parameter $m=1$.  If $\Omega$ is not an isotropic subspace then it can be written compactly as
	\begin{equation}\label{CNCSets}
		\Omega=\bigcup\limits_{k=1}^\xi\langle p_k,\tilde{I}\rangle
	\end{equation}
	where $\xi\ge2$, the cosets $p_1+\tilde{I},\dots,p_\xi+\tilde{I}$ are distinct and $q_1,\dots,q_\mu$ are in the cosets $p_1+\tilde{I},\dots,p_\xi+\tilde{I}$.  Note that in this form, there can be no element $p_j$ which commutes with all of $p_1,\dots,p_\xi$ because $\tilde{I}$ is defined to contain all such elements.  Now we consider the possible commutation relations that $p_1,\dots,p_\xi$ can have if $\Omega$ is non-contextual.\medskip
	
	The Mermin square is generated by products of commuting pairs of the two qubit Pauli operators $\{X_1,X_2,Z_1,Z_2\}$.  This is a contextual set.  Therefore, any set which is closed under inference and contains four elements $p_1,p_2,p_3,p_4$ with the commutation relations like those of $\{X_1,X_2,Z_1,Z_2\}$:
	\begin{equation}\label{CR1}
	\begin{split}
	[p_1,p_2]&=[p_1,p_4]=[p_2,p_3]=[p_3,p_4]=0\\
	[p_1,p_3]&=[p_2,p_4]=1
	\end{split}
	\end{equation}
	will necessarily contain the full Mermin square and therefore be contextual.
	
	Another sufficient condition for a closed under inference set to be contextual is that it contains four elements with the commutation relations
	\begin{equation}\label{CR2}
	\begin{split}
	[p_1,p_2] & =[p_2,p_3]=[p_3,p_4]=0\\
	[p_1,p_3] & =[p_1,p_4]=[p_2,p_4]=1.
	\end{split}
	\end{equation}
	The reason is that since the set is closed under inference, it will necessarily contain the elements $p_1+p_2$ and $p_3+p_4$, and the elements $p_1,p_1+p_2,p_3+p_4,p_4$ have the commutation relations of Eq.~(\ref{CR1}).  Thus, it must contain a Mermin square.
	
	A similar argument shows that another sufficient condition for a closed under inference set to be contextual is that it contains four elements with the commutation relations
	\begin{equation}\label{CR3}
	\begin{split}
	[p_1,p_2] & =[p_2,p_3]=0\\
	[p_1,p_3] & =[p_1,p_4]=[p_2,p_4]=[p_3,p_4]=1.
	\end{split}
	\end{equation}
	In this case, since the set is closed under inference, it must also contain the elements $p_1+p_2$ and $p_2+p_3$ and the elements $p_1+p_2,p_2,p_2+p_3,p_4$ have the commutation relations of Eq.~(\ref{CR1}).\medskip
	
	To determine the possible commutation relations of the elements $p_1,\dots,p_\xi$, we will look at their commutativity graph $\mathcal{G}$.  That is the undirected graph with a vertex for each of $p_1,\dots,p_\xi$ and an edge connecting each pair of commuting vertices.  Since $\Omega$ is non-contextual, the commutation relations of Eq.~(\ref{CR1}), Eq.~(\ref{CR2}) and Eq.~(\ref{CR3}) provide restrictions on the possible commutation relations of the elements $p_1,\dots,p_\xi$ of $\Omega$.  In terms of the commutativity graph $\mathcal{G}$, these are forbidden induced subgraphs\footnote{An induced subgraph of a graph is the graph obtained by taking a subset of the vertices of the original graph and all of the edges connecting pairs of vertices in the subset}.
	
	The restriction of Eq.~(\ref{CR1}) says that $\mathcal{G}$ cannot have a four vertex chordless cycle ($C_4$) as an induced subgraph and the restriction from Eq.~(\ref{CR2}) says that $\mathcal{G}$ cannot have a four vertex path ($P_4$) as an induced subgraph.  These two forbidden induced subgraphs characterize the trivially perfect graphs~\cite{Golumbic77}.  I.e. $\mathcal{G}$ must be a trivially perfect graph.
	
\begin{figure}
\begin{center}	
\begin{tabular}{ccccc}
	\includegraphics[width=1.5cm]{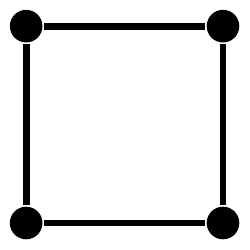} &\hspace*{4mm}&
	\includegraphics[width=1.5cm]{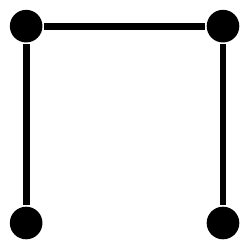} &\hspace*{4mm}&
	\includegraphics[width=1.5cm]{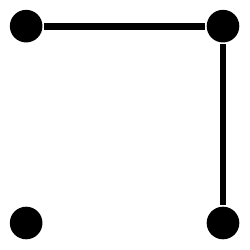} \\
	$C_4$ && $P_4$ && $K_1\cup P_3$
\end{tabular}
\caption{\label{ExclSG}Forbidden induced subgraphs of the commutativity graph, resulting from Mermin's square (also see \cite{KirbyLove}).}
\end{center}
\end{figure}
	
Connected trivially perfect graphs have the the property that they contain a universal vertex \cite{Golumbic77}\footnote{A universal vertex is a vertex that is adjacent to every other vertex in the graph.}.  If the commutativity graph $\mathcal{G}$ were connected then there would be an element $p_j$ which commutes with all other elements of $\{p_1,\dots,p_\xi\}$.  This is also forbidden.  Therefore, the graph $\mathcal{G}$ is disconnected.
	
Given that $\mathcal{G}$ is disconnected, Eq.~(\ref{CR3}) provides another restriction.  Namely that each connected component of $\mathcal{G}$ cannot have a three vertex path ($P_3$) as an induced subgraph.  I.e. each connected component of $\mathcal{G}$ is a clique.
	
This means we can partition the elements $\{p_1,\dots,p_\xi\}$ into disjoint subsets	
\begin{equation}\label{setPart}
\begin{array}{rcl}
\{p_1,\dots,p_\xi\} &= &\{p_{1,1},p_{1,2},\dots,p_{1,\xi_1}\}\cup \\
&& \{p_{2,1},p_{2,2},\dots,p_{2,\xi_2}\} \cup \cdots \cup \\
&& \{p_{\pi,1},p_{\pi,2},\dots,p_{\pi,\xi_\pi}\}
\end{array}
\end{equation}
where two elements commute if and only if they are in the same subset in the partition.  Since the set $\{p_1,\dots,p_\xi\}$ is closed under inference, each subset in the partition must be closed under inference.  Now suppose a subset in the partition contained at least two elements.  Then since the subset is closed under inference it must also contain their sum.  But each of the two elements anticommutes with the elements of all other subsets in the partition so their sum must commute with the elements of all other subsets in the partition.  This is a contradiction.  Therefore, each subset in the partition contains a single element.  Thus, the elements $\{p_1,\dots,p_\xi\}$ of Eq.~(\ref{CNCSets}) pair-wise anticommute.
	
Maximal cnc sets are sets of the form Eq.~(\ref{CNCSets}) for which $\xi$ is maximal for a given isotropic subspace $\tilde{I}$.  If the isotropic subspace $\tilde{I}$ has dimension $n-m$ where $n$ is the number of qubits and $1\le m\le n$, then the pair-wise anticommuting elements $p_k$ which complete the set are elements of the symplectic complement $\tilde{I}^{\perp}$.  This is a $m$ dimensional symplectic subspace, therefore the maximal value of $\xi$ is the largest number of pair-wise anticommuting Pauli operators on $m$ qubits.  The largest sets of pair-wise anticommuting Pauli operators on $m$ qubits have $2m+1$ elements.  This can be seen as follows. Consider the elements $a_k\in E$ given by $T_{a_k}=C_k$, with $C_k$ defined in Eq.~(\ref{Majorana}).
The set $\lbrace a_k|\; 1\leq k\leq 2m \rbrace$ consists of pairwise anticommuting elements. There is an element $c$, with $T_c= C_{2m+1}$, cf. Eq.~(\ref{MajoB}) that anticommutes with each one of the elements in this set. It is the only element in $E$ to do so, since the set of equations
\begin{equation}\label{eq:com-rel}
[c,a_k]=1\;\;\;\; 1\leq k\leq 2m
\end{equation}
has a unique solution. Therefore together with this element we can construct a set of size $2m+1$. 

We would like to show  any other set of pairwise anticommuting elements whose size is $2m$ can be mapped bijectively to the set we constructed. Suppose  $\lbrace \tilde a_k|\; 1\leq k\leq 2m \rbrace$ is such a set. By Witt's lemma \cite[\S 20]{Asch} the function that sends $ a_k$ to $\tilde a_k$  extends to a linear map $f:E\to E$ that satisfies $[f(v),f(w)]=[v,w]$ for all $v,w\in E$ (symplectic transformation). Therefore there is a unique element that anticommutes with all the $\tilde a_k$, and it is given by $f(c)$. In particular, $2m+1$ is the maximal number.

To complete the proof we must show that maximal sets of pair-wise anticommuting elements on $m$ qubits with size less than $2m+1$ do not lead to maximal cnc sets.  To see this note that by Witt's lemma, for any maximal anticommuting set of size $2m'+1$ ($m'<m$), there is a bijection $f:E\rightarrow E$ which maps the set to one of the form Eq.~(\ref{Majorana},\ref{MajoB}).  Therefore, we can find $m-m'$ independent elements which commute with the set.  For example, if $g_1, g_2,\dots,g_{m-m'}$ are the vectors corresponding to Pauli operators $X_{m'+1},X_{m'+2},\dots,X_{m}$, then we could take $f^{-1}(g_1),f^{-1}(g_2),\dots,f^{-1}(g_{m-m'})$.  Therefore, the $n-m$ dimensional isotropic subspace can be extended to one with dimension $n-m'$.

This completes the proof.  Therefore, all maximal cnc sets have the form Eq.~(\ref{manyI}), with $\xi=2m+1$. $\Box$}\medskip

A result equivalent to the characterization of Eq.~(\ref{setPart}) is given in Theorem 3 of~\cite{KirbyLove}.

Tensor products of phase point operators are not typically phase point operators. Consider, for example, two phase point operators with $m=2$ and $n\geq 2$. Their tensor product does not appear in the classification provided by Theorem~\ref{nms}, as the commutativity graph shows. Physically, such tensor products are not closed under inference, violating Def.~\ref{PhaSpa}. Upon closure, they cease to be non-contextual as they then contain a Mermin square. Hence the closures also violate Def.~\ref{PhaSpa}.

But there is an exception. If one of the two phase point operators in the tensor product corresponds to an isotropic subspace, i.e., has $m=0$, then the tensor product {\em{is}} a valid phase point operator. See Appendix~\ref{LTpr} for details.

\subsection{Relation to the stabilizer formalism}\label{Exa}

The purpose of this section is to describe the relation between positive representability by the quasiprobability distribution $W$ and qubit stabilizer states. We demonstrate that, for all $n$, the set of positively $W$-representable states contains the stabilizer mixtures as a strict subset. This is the content of Lemma~\ref{Exa23} below. The lemma is based on two examples.\medskip

{\em{Example 4.}} Be $|\text{stab}\rangle$ an $n$-qubit stabilizer state, with isotropic subspace $\tilde{I} \subset E$ corresponding to its stabilizer. Then, it is easily verified that $\tilde{I}$ is non-contextual and closed under inference. Namely, $\tilde{I}$ is of form Eq.~(\ref{manyI}), with $m=0$, $\xi=1$.  \medskip

The next example generalizes Example 1 to $n$-qubit states.\medskip

{\em{Example 5.}} Every $n$-qubit state of the form $\Psi = \rho_1 \otimes |\text{stab}\rangle \langle \text{stab}|_{2,..,n}$, with $\rho$ a general one-qubit state and $|\text{stab}\rangle$ an $n-1$-qubit pure stabilizer state, is positively representable. 

To prove this statement, for any number $n$ of qubits, consider an isotropic subspace $\tilde{I}\subset E$ of rank $n-1$ representing the stabilizer state $|\text{stab}\rangle_{2,..,n}$, and three elements $x,y,z\in E$, such that $T_x=X_1$, $T_y=Y_1$ and $T_z=Z_1$. Define the three isotropic subspaces $I_x,I_y,I_z \subset E$,
$$
I_x = \langle x,\tilde{I}\rangle,\; I_y = \langle y,\tilde{I}\rangle,\; I_z = \langle z,\tilde{I}\rangle,
$$
and  $\Omega_{xyz}:= I_x \cup I_y \cup I_z$. $\Omega_{xyz}$ is of form Eq.~(\ref{manyI}), with $m=1$, $\xi=3$ and $n\geq 2$, hence cnc by Lemma~\ref{L4}.

We now apply this result to the state $\Psi= \rho_1 \otimes |\text{stab}\rangle \langle \text{stab}|_{2,..,n}$ above. We can write the constitutents as $\rho=\sum_{\gamma_0} W_\rho(\Omega_0,\gamma_0) A_{\Omega_0}^{\gamma_0}$, with $W_\rho\geq 0$ (cf. Example 1), and $|\text{stab}\rangle \langle \text{stab}|=A_{\tilde{I}}^{\tilde{\gamma}}$ (cf. Example 2). We observe that 
$$
A_{\Omega_0}^{\gamma_0}\otimes A_{\tilde{I}}^{\tilde{\gamma}}=A_{\Omega_{xyz}}^\gamma,
$$
with $\gamma:=\gamma_0(a|_1)+\tilde{\gamma}(a|_{2,..,n}) \mod 2$. To see this, recall that $\Omega_0=\{0,x,y,x\}$, and note that the set $\Omega_{xyz}$ can also be written as $\Omega_{xyz} = \tilde{I} \cup (\tilde{I}+x)  \cup (\tilde{I}+y) \cup (\tilde{I}+z)$, where the coset $\tilde{I}+x:=\{a+x,\; \forall a\in \tilde{I}\}$, etc.
Thus, $\Psi=\sum_{\gamma_0} W_\rho(\Omega_0,\gamma_0) A_{\Omega_{xyz}}^\gamma$. Since $W_\rho\geq 0$ by Example 1, the states $\Psi$ are all positively representable. Yet not all these states are mixtures of stabilizer states. Stabilizer mixedness is preserved under partial trace. Now assume that $\Psi$ is a stabilizer mixture for all $\rho$. Then $\text{Tr}_{2..n}\Psi =\rho_1$ is also a stabilizer mixture. Contradiction.\medskip

We cast the combined conclusion of Examples 4 and 5 as a Lemma.
\begin{Lemma}\label{Exa23}
For all $n\in \mathbb{N}$, all mixtures of $n$-qubit stabilizer states are positively representable, and furthermore there exist positively representable states that are not mixtures of stabilizer states.
\end{Lemma}

\section{Quantum mechanical rules for state update under measurement}\label{QUP}

In the previous sections we have analyzed the generalized phase space ${\cal{V}}$ on which the quasiprobability function $W$ is defined. We now turn to dynamics.

For our setting of QCM this concerns evolution under the free operations, i.e., the Clifford unitaries and Pauli measurements. As already noted in \cite{ReWi} and \cite{QuWi}, the situation simplifies even further. If the goal is to sample from the joint probability distribution of measurement outcomes---which is the case in quantum computation---then only the update under Pauli measurements needs to be considered. 

The Clifford unitaries can be propagated forward in time, thereby conjugating the Pauli measurements into other such measurements, past the final measurement and then discarded. (This redundancy notwithstanding, we will visit the update of $W$ under Clifford unitaries in Section~\ref{PhaSpaStruct}, where we prove covariance.) The main results of this section are Theorem~\ref{CPM} and Lemma~\ref{Q_up}. 

\begin{Theorem}\label{CPM}
For any $n\in \mathbb{N}$, the set ${\cal{P}}_n$ of positively representable $n$-qubit quantum states is closed under Pauli measurement.
\end{Theorem}

To describe the dynamics under measurement, we need to set up some further notation. For every set $\Omega$ we introduce the derived set $\Omega \times a$. Denoting $\text{Comm}(a):=\{ b \in E| [a,b]=0\}$ and $\Omega_a := \Omega \cap \text{Comm}(a)$,
\begin{equation}
	\label{Def_OmB}
	\Omega\times a := \Omega_a \cup \{a+b|\, b\in \Omega_a\},\;\;\; \forall a\not \in \Omega.
\end{equation}
Likewise, we define an update on functions $\gamma$ invoking the measurement outcome $s_a$ of an observable $T_a$, namely $(\cdot)\times s_a: \left(\gamma: \Omega \longrightarrow \mathbb{Z}_2\right) \mapsto  \left(\gamma\times s_a: \Omega\times a \longrightarrow \mathbb{Z}_2\right)$. We define this update only for $(\Omega,\gamma)\in {\cal{V}}$, and only for $a\not\in \Omega$ \footnote{The definitions of $\Omega\times a$ and $\gamma\times s_a$ can without modification be extended to $a\in \Omega$. However, in that case the function values $\gamma\times s_a(b)$ can be determined both through Eq.~(\ref{gammaA}) and (\ref{gammaB}), and we need to check consistency. These inferences are indeed consistent, as a consequence of Eq.~(\ref{dgb}). Since we do not need the case of $a\in \Omega$ subsequently, we skip the details of the argument.}. The updated function $\gamma \times s_a: \Omega\times a \longrightarrow \mathbb{Z}_2$ is given by
\begin{subequations}\label{gammaUp}
	\begin{align}\label{gammaA}
	\gamma\times s_a(b) &:= \gamma(b),& \forall b \in \Omega_a,\\
	\label{gammaB}
	\gamma \times s_a(b) &:= \gamma(a+ b) + s_a+\beta(a,b),& \forall a+ b \in \Omega_a.
	\end{align}
\end{subequations} 
The rules of Eq.~(\ref{gammaUp}) are used to formulate the update rule for phase point operators of Eq.~(\ref{PPO2}) under Pauli measurement.

{\em{Remark.}} Update rules similar to Eq.~(\ref{gammaUp}) have been used previously \cite{Lilly} to construct a $\psi$-epistemic model of the multi-qubit stabilizer formalism. Those rules update the value assignments in the same way but are applied under different conditions. Specifically, the update in \cite{Lilly} does not refer to general sets $\Omega$ satisfying the conditions of Def~\ref{PhaSpa}.\medskip

\begin{Lemma}\label{Q_up} Denote the projectors $P_a(s_a):= (I +(-1)^{s_a}T_a)/2$, and be $A_\Omega^\gamma$ a phase point operator defined through Eq.~(\ref{PPO2}), with $(\Omega,\gamma)\in {\cal{V}}$ satisfying the conditions of Definition~\ref{PhaSpa}. Then, the effect of a measurement of the Pauli observable $T_a$ with outcome $s_a$ on $A_\Omega^\gamma$ is
\begin{subequations}
\label{A_up}
\begin{align}
\label{A_up_in}
P_a(s_a) A_\Omega^\gamma P_a(s_a)  &= \delta_{s_a,\gamma(a)} \frac{A_\Omega^\gamma + A_\Omega^{\gamma+ [a,\cdot]}}{2},&\text{if}\; a\in \Omega,\\
\label{A_up_out}
P_a(s_a) A_\Omega^\gamma P_a(s_a) &= \frac{1}{2} A_{\Omega \times a}^{\gamma \times s_a}, &\text{if}\; a\not\in \Omega.
\end{align}
\end{subequations}
\end{Lemma}
\begin{figure}
\begin{center}
\begin{tabular}{lc}
(a) &\\
& \includegraphics[width=6.5cm]{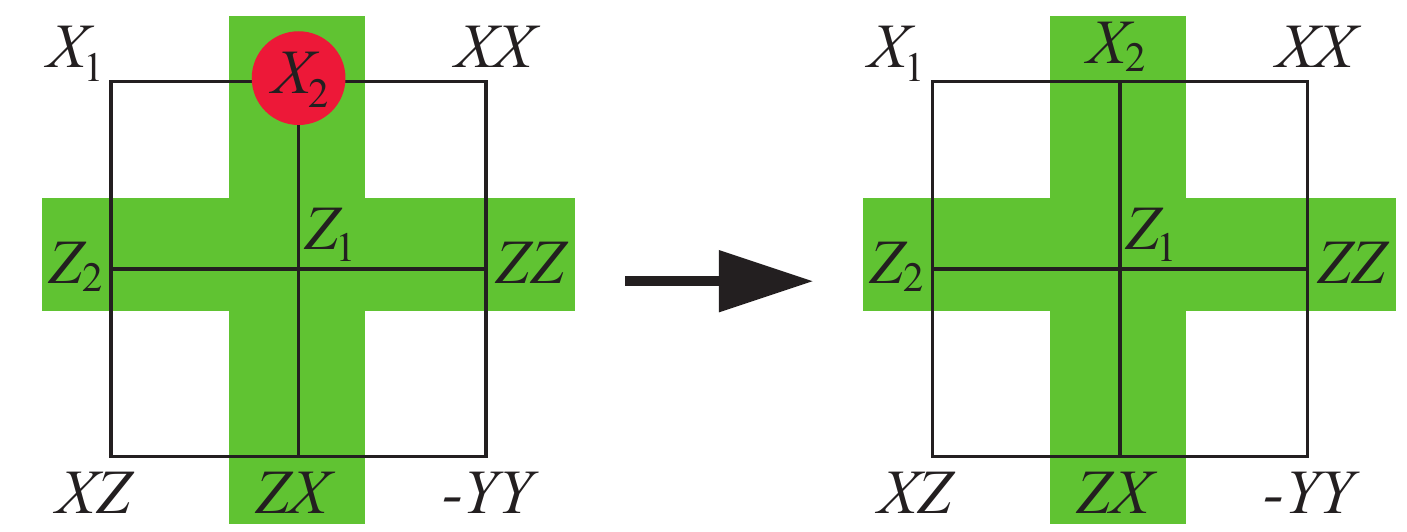} \\

(b) &\\
&  \includegraphics[width=6.5cm]{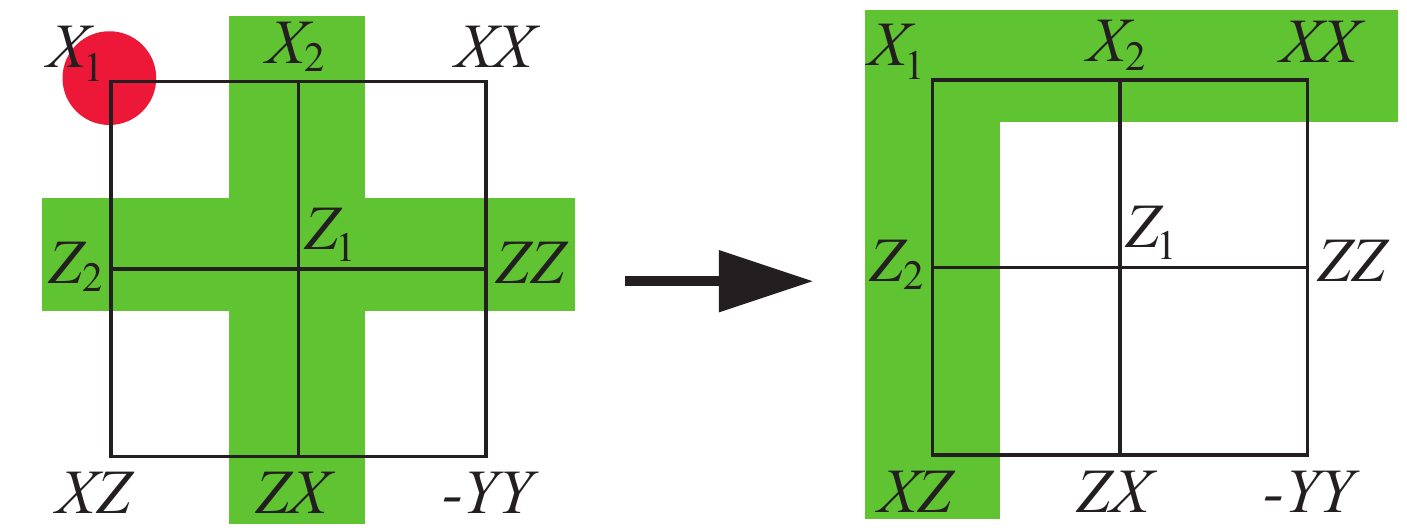}
\end{tabular}
\caption{\label{M_up}Update of a cnc set $\Omega$ in Mermin's square, under two Pauli measurements. (a) The measured observable $X_2$ is such that $a(X_2) \in \Omega$, hence the update proceeds by Eq.~(\ref{A_up_in}). (b) The measured observable $X_1$ is such that $a(X_1) \not \in \Omega$; hence the update proceeds by Eq.~(\ref{A_up_out}).} 
\end{center}
\end{figure}

{\em{Example 2, continued.}} Eq.~(\ref{A_up}) entails the update of both the sets $\Omega$ and the functions $\gamma$. Here we only consider the former. Fig.~\ref{M_up} displays the update of the set $\Omega$ shown in Fig.~\ref{MermOm} a, under the measurement of (a) the observable $X_2$, with $a(X_2)\in \Omega$, and (b) the observable $X_1$, with $a(X_1) \not\in \Omega$. \medskip 

In preparation for the proof of Lemma~\ref{M_up} it is useful to state two relations of the function $\beta$ for $d=2$. With the definition Eq.~(\ref{3T}) of $\beta$ and Eq.~(\ref{NormCond}), the operator identities $T_aT_a=I$ and $T_b I = T_b$ imply that
\begin{equation}\label{beta_constr2}	
\beta(a,a)=\beta(a,0)=\gamma(0),\;\; \forall a\in E.
\end{equation}
Furthermore, evaluating $d\beta (a,a,0)=0$ (see Eqs.~(\ref{beta_constr}) and (\ref{Def_deb})), and using Eq.~(\ref{beta_constr2}) yields
\begin{equation}\label{beta2}
\beta(a,b) = \beta(a,a+b),\;\; \forall a,b.
\end{equation}
To prove Lemma~\ref{Q_up} we also need the following result.
\begin{Lemma}\label{L1}
If $\Omega \subset E$ is non-contextual and closed under inference, then so is $\Omega_a$, for all $a\in E$.
\end{Lemma}
{\em{Proof of Lemma~\ref{L1}.}} First consider closure. Assume that $c,d\in \Omega_a$ and $[c,d]=0$. Then, $c,d\in \Omega$, and also $c+ d \in \Omega$, since $\Omega$ is closed by assumption. Further, $[c,a]=[d,a]=0$ implies $[c+ d,a]=0$, and hence $c+ d\in \Omega_a$. $\Omega_a$ is thus closed.

Now consider non-contextuality. Since $\Omega$ is non-contextual, there exists a function $\gamma$ such that $d\gamma = \beta$ on $\Omega$. Since $\Omega_a$ is closed, $\beta$ can be properly restricted to ${\cal{C}}(\Omega_a)$, and so can $\gamma$. Hence, $d \gamma|_{{\cal{C}}(\Omega_a)}=\beta|_{{\cal{C}}(\Omega_a)}$. Thus, $\Omega_a$ is non-contextual. $\Box$\medskip

{\em{Proof of Lemma~\ref{Q_up}.}} Under the measurement of $T_a$ with outcome $s_a\in \mathbb{Z}_2$ we have
\begin{equation}\label{comStep}
\begin{array}{l}
\displaystyle{\frac{I + (-1)^{s_a}T_a}{2} A_\Omega^\gamma \frac{I + (-1)^{s_a}T_a}{2}} =\vspace{2mm}\\ 
=\displaystyle{\frac{I + (-1)^{s_a}T_a}{2} \frac{1}{2^n} \sum_{b \in \Omega_a}(-1)^{\gamma(b)} T_b} \vspace{2mm}\\
=\displaystyle{ \frac{1}{2\cdot2^n} \sum_{b \in \Omega_a}(-1)^{\gamma(b)} T_b+ \frac{(-1)^{s_a}}{2\cdot2^n} \sum_{b \in \Omega_a}(-1)^{\gamma(b)} T_aT_b}.
\end{array}
\end{equation}
From hereon we need to distinguish two cases, $a\in \Omega$ and $a\not\in \Omega$.\smallskip

{\em{Case I: $a\in \Omega$.}} 
Focusing on the second term in the expansion Eq.~(\ref{comStep}),
$$
\begin{array}{l}
\displaystyle{(-1)^{s_a}\sum_{b \in \Omega_a}(-1)^{\gamma(b)} T_aT_b} = \\
\hspace*{15mm}=\displaystyle{(-1)^{s_a}\sum_{b \in \Omega_a}(-1)^{\gamma(b)+\beta(a,b)} T_{a+ b}}\\
\hspace*{15mm}= \displaystyle{(-1)^{s_a+\gamma(a)}\sum_{b \in \Omega_a}(-1)^{\gamma(a+ b)} T_{a+ b}}\\
\hspace*{15mm}= \displaystyle{(-1)^{s_a+\gamma(a)}\sum_{a + b \in \Omega_a}(-1)^{\gamma(a+ b)} T_{a+ b}}\\
\hspace*{15mm}= \displaystyle{(-1)^{s_a+\gamma(a)}\sum_{b \in \Omega_a}(-1)^{\gamma(b)} T_{b}}.
\end{array}
$$
Therein, in the first line we have used Eq.~(\ref{3T}), in the second line Eq.~(\ref{dgb}), in the third line the completeness of $\Omega_a$ under inference (Lemma~\ref{L1}), and the fourth line is just a relabeling of the elements in $\Omega_a$. Inserting this result in the above expansion Eq.~(\ref{comStep}), we find
\begin{equation}\label{AltUp}
P_a(s_a) A_\Omega^\gamma P_a(s_a) = \delta_{s_a,\gamma(a)} \frac{1}{2^n} \sum_{b \in \Omega_a}(-1)^{\gamma(b)} T_{b},
\end{equation}
and Eq.~(\ref{A_up_in}) follows.

{\em{Case II: $a \not\in \Omega$.}}  Substituting $b \longrightarrow a+b$ in Eq.~(\ref{gammaB}) gives $\gamma\times s_a(a+b)=\gamma(b) + s_a+\beta(a,a+b)$, for  $b\in \Omega_a$. With Eq.~(\ref{beta2}) we obtain
\begin{equation}
\label{gammaB1}
\gamma\times s_a(a+b)=\gamma(b) + s_a+\beta(a,b),\;\;\forall b \in \Omega_a.
\end{equation}
With this, we now look at the second term in the expansion Eq.~(\ref{comStep}),
$$
\begin{array}{l}
\displaystyle{(-1)^{s_a}\sum_{b \in \Omega_a}(-1)^{\gamma(b)} T_aT_b} =\\
\hspace*{15mm}=  \displaystyle{(-1)^{s_a}\sum_{b \in \Omega_a}(-1)^{\gamma(b)+\beta(a,b)} T_{a+ b}}\\
\hspace*{15mm}= \displaystyle{\sum_{b \in \Omega_a}(-1)^{\gamma\times s_a(a+ b)} T_{a+ b}}.
\end{array}
$$
The first line above follows with Eq.~(\ref{3T}), and the second with Eq.~(\ref{gammaB1}).

Considering the first term in the expansion Eq.~(\ref{comStep}), with Eq.~(\ref{gammaA}) we have
$$
 \sum_{b \in \Omega_a}(-1)^{\gamma(b)} T_b =  \sum_{b \in \Omega_a}(-1)^{\gamma\times s_a(b)} T_b
$$
Inserting the above expressions for the two terms in Eq.~(\ref{comStep}), and using the definition Eq.~(\ref{Def_OmB}) of $\Omega \times a$, we obtain Eq.~(\ref{A_up_out}). $\Box$
\medskip

We have so far shown how the phase point operators can be updated under measurement once. We still need to show that this update can be iterated. This requires that the new phase point operators appearing on the r.h.s. of Eq.~(\ref{A_up}) satisfy the consistency constraints of Definition~\ref{PhaSpa}.

\begin{Lemma}\label{Perpet}
If $(\Omega, \gamma) \in {\cal{V}}$ then $(\Omega, \gamma + [a,\cdot]) \in {\cal{V}}$, for all $a\in \Omega$, and $(\Omega\times a, \gamma\times s_a) \in {\cal{V}}$, for all $a\not\in \Omega$ and $s_a \in \mathbb{Z}_2$.
\end{Lemma}
The proof of Lemma~\ref{Perpet} is given in Appendix~\ref{PerpetProof}.\medskip

{\em{Proof of Theorem~\ref{CPM}.}} Consider a state $\rho \in {\cal{P}}_n$, and a measurement of the Pauli observable $T_a$ on it. Assume that the measurement outcome $s_a$ can occur, $p_a(s_a):=\text{Tr}(P_a(s_a)\rho)>0$. We have to show that under these conditions, the post-measurement state 
$$
\rho' = \frac{P_a(s_a) \rho P_a(s_a)}{p_a(s_a)}
$$
is also contained in the set ${\cal{P}}_n$.

Denote $\overline{\delta}_{a\in \Omega} := 1 - \delta_{a\in\Omega}$. Then, with Lemma~\ref{Q_up} and the state expansion Eq.~(\ref{Wigner}) of $\rho$, we have
\begin{widetext}
\begin{equation}\label{PMS}
\rho' = \sum\limits_{(\Omega,\gamma) \in {\cal{V}}} \frac{W_\rho(\Omega,\gamma)}{p_a(s_a)}\Big( \delta_{a\in\Omega}\delta_{s_a,\gamma(a)}\frac{A_\Omega^\gamma + A_\Omega^{\gamma+[a,\cdot]}}{2}+\frac{1}{2} \overline{\delta}_{a\in\Omega} A_{\Omega\times a}^{\gamma \times s_a} \Big).
\end{equation}
Thus, $\rho'$ can be represented by a quasiprobability distribution $W_{\rho'}$ with elements
\begin{equation}\label{W_elements}
W_{\rho'}(\Omega',\gamma') = \!\!\sum_{(\Omega,\gamma) \in {\cal{V}}} \!\!\frac{W_\rho(\Omega,\gamma)}{2 p_a(s_a)} \left(\delta_{a\in\Omega} \delta_{s_a,\gamma(a)} \left(\delta_{(\Omega',\gamma'), (\Omega,\gamma)}  +\delta_{(\Omega',\gamma'),(\Omega, \gamma+[a,\cdot])} \right) + \overline{\delta}_{a\in\Omega} \delta_{(\Omega',\gamma'),(\Omega \times a, \gamma\times s_a)} \right).
\end{equation}
\end{widetext}
The $W_{\rho'}(\Omega',\gamma')$ are thus linear combinations of $W_\rho(\Omega,\gamma)$ with non-negative coefficients (0 or $1/2p_a(s_a)$). Since the $W_\rho(\Omega,\gamma)$ are non-negative by assumption, it follows that $W_{\rho'}(\Omega',\gamma') \geq 0$, for all $(\Omega',\gamma') \in {\cal{V}}$.
$\Box$

\section{Classical simulation for $W_\rho\geq 0$}\label{CEsim}

\subsection{Simulation algorithm}\label{Calgo}

We now turn to the question of how hard it is to classically simulate the outcome statistics for a sequence of Pauli measurements on an initial quantum state. In this regard, we show that if the initial quantum state is positively represented and the corresponding probability distribution $W$ can be efficiently sampled from, then the statistics of the measurement outcomes can be efficiently simulated.

The classical simulation procedure in Table~\ref{SimAlg} describes weak simulation \cite{VdN1}--\cite{VdN2}, i.e., it outputs one sample from the joint probability distribution $p(s_{a_1}, s_{a_2},..,s_{a_N})$ of outcomes corresponding to a sequence of measurements of Pauli operators $T_{a_1},T_{a_2}, .. ,T_{a_N}$ ($T_{a_1}$ is measured first, $T_{a_N}$ last). If more than one sample are desired, the procedure is just repeated. We note that the measurement can be adaptive. I.e., it is not necessary for the simulation algorithm that a measurement sequence is committed to at the beginning. As a special case of this, the measured observables may depend on earlier measurement outcomes.

\begin{table}
\begin{center}
\textbf{Classical simulation algorithm}\vspace{2mm}

\fbox{\parbox{7.5cm}{
\hspace*{-3mm}\parbox{7cm}{
\begin{enumerate}
\item{Draw a sample $(\Omega,\gamma) \in {\cal{V}}$ according to the probability distribution $W_\rho$ representing the initial quantum state $\rho$.}
\item{For the observables $T_{a_1}, T_{a_2}, .. , T_{a_N}$ measured in this sequence, repeat the following steps. 

For the $i$-th measurement, set $a:=a_i$. 

If $a \in \Omega$ then $\Omega$ is unchanged. Output the value $s_a=\gamma(a)$. Flip a coin.
$$
\begin{array}{ll}
\text{if ``heads''} & \text{then } \gamma \longrightarrow \gamma,\\
\text{if ``tails''} & \text{then } \gamma \longrightarrow \gamma+ [a,\cdot].
\end{array}
$$
If $a \not \in \Omega$ then $\Omega\longrightarrow \Omega \times a$. Flip a coin.
$$
\begin{array}{ll}
\text{if ``heads''} & \text{then } s_a=0,\\
\text{if ``tails''} & \text{then } s_a=1.
\end{array}
$$
Output this value $s_a$. Update $\gamma \longrightarrow \gamma\times s_a$, through Eq.~(\ref{gammaUp}).}
\end{enumerate}}}}
\caption{\label{SimAlg}Classical simulation algorithm for sampling from the joint probability distribution of a sequence of Pauli measurements on a positively represented initial quantum state.}
\end{center}
\end{table}
We have the following result.
\begin{Theorem}\label{T1}
If for an initial quantum state $\rho$ it holds that $W_\rho\geq 0$ and furthermore $W_\rho$ can be efficiently sampled from, then the output distribution of all sequences of Pauli measurements, possibly interspersed with Clifford gates, on $\rho$ can be classically efficiently sampled from. 
\end{Theorem}
As a first application of Theorem~\ref{T1}, we return to Example 2, Mermin's square.\smallskip

\begin{table}
\begin{center}
\begin{tabular}{|l|r|r|r|}
\hline
$m$ & $0$ & $1$ & $\{1,2\}$\\ \hline\hline
2 rebits  & 24 & 72 & 120 \\ \hline
2 qubits & 60 & 240 & 432\\ \hline
\end{tabular}
\caption{\label{Mem}Number of points in phase space as a function of $\{m\}$.}
\end{center}
\end{table}

{\em{Example 2 continued.}} How much memory capacity is needed to classically simulate measurements of the observables in Mermin's square? We first turn to the state-independent case, which was previously discussed in \cite{MemCo}. The task is to devise a classical algorithm that outputs an outcome sequence for any given sequence of Pauli measurements, which can occur according to quantum mechanics. The measurement sequence can be of any length and the measurements therein may be commuting or anti-commuting. In \cite{MemCo}, a lower bound on the memory cost of any such simulation was established, $\log_2 24$ bits; and a specific model was constructed that attains it.

The classical simulation algorithm of Table~\ref{SimAlg} also saturates this limit. To show this, we use as cnc sets $\Omega$ the six maximal isotropic subspaces of two rebits, cf. Fig.~\ref{MermOm} b. This set of sets $\Omega$ is closed under update by Pauli measurement, as described by Eq.~(\ref{A_up}). For each such set $\Omega$, each value assignment $\gamma$ is specified by two evaluations (the other evaluations then follow via Eq.~(\ref{dgb})). There are thus four functions $\gamma$ for each cnc set $\Omega$, hence 24 combinations in total, which is the same as in \cite{MemCo}.

We now turn to the state-dependent version of the problem. How much memory is needed to sample from the correct outcome statistics for arbitrary measurement sequences, for any two-rebit state $\rho$ with $W_\rho\geq 0$, and given the capability to sample from $W_\rho$? This problem is harder than the former: Not only must the sequence of outcomes be internally consistent for all measurement sequences, but also it needs to represent the state $\rho$. 

Memory cost now depends on the state $\rho$. If $\rho$ is a mixture of stabilizer states, i.e., the sets $\Omega$ can be limited to $m=0$, then the classical simulation algorithm of Table~\ref{SimAlg} can still run on $\log_2 24\approx 4.59$ bits. 

If sets $\Omega$ with $m=1$ are included in the expansion, then more two-rebit states $\rho$ can be positively represented (among them, for example, $|T\rangle_1 \otimes |T\rangle_2$) but on the other hand, memory consumption goes up. For $m=1$, there are $3^2\times 2^3$ pairs $(\Omega,\gamma)$, cf. Fig.~\ref{MermOm}a. Hence the memory consumption for configurations with $m=1$ is  $\log_2 72\approx 6.17$ bits. (Note that the sets $\Omega$ for $m=0$ are not maximal. If sets with $m=1$ are included, then sets with $m=0$ can be omitted without loss.) The size $|{\cal{V}}(\{m\})|$ of the phase space vs. the maximum value of $m$ is displayed in Table~\ref{Mem}. The memory cost is $\log_2 |{\cal{V}}(\{m\})|$. The volume fraction of positively representable two-rebit and two-qubit states is displayed in Table.~\ref{Vol}, for various sets $\{m\}$.

\begin{table}
\begin{center}
Two rebits\vspace{1mm}\\

\begin{tabular}{|l|r|r|r||r|}
\hline
$m$ & $0$ & $1$ & $\{1,2\}$ & hy.oct.\\ \hline\hline
$V_+/V$ [pure] & 0 & 1 & 1 & 0\\ \hline
$V_+/V$ [mixed] & 0.144  & 1 & 1 & 0.924\\ \hline
\end{tabular}\vspace{5mm}\\

Two qubits\vspace{1mm}\\

\begin{tabular}{|l|r|r|r||r|}
\hline
$m$ & $0$ & $1$ & $\{1,2\}$ & hy.oct.\\ \hline\hline
$V_+/V$ [pure]  & 0 & 0.980  & 0.980 & 0\\ \hline
$V_+/V$ [mixed]  & 0.009 & 1 & 1 & 0.568\\ \hline
\end{tabular}
\caption{\label{Vol}Volume fraction of state space filled by the positively representable states, as a function of $\{m\}$; (top) two rebits, (bottom) two qubits. The volume fraction $V_+/V$ was obtained numerically, by sampling $10^6$ random states according to the Fubini-Study measure for pure states (second row) and the Hilbert-Schmidt measure for mixed states (third row). The first column, $m=0$, describes mixtures of stabilizer states, and the last column hyper-octahedral states \cite{Rall} for comparison.}
\end{center}
\end{table}

\subsection{Correctness and efficiency of the classical simulation}\label{CandE}

In preparation for the proof of correctness of the classical simulation algorithm, we introduce the following notation. Given a probability distribution $W_\rho$, there are two objects that the classical simulation algorithm needs to reproduce correctly, namely the probability $p_a(s_a)$ for the outcomes $s_a \in \mathbb{Z}_2$ of the measurement of any Pauli observable $T_a$, and the post-measurement state $\rho'$.  There are two ways of obtaining these quantities, a quantum-mechanical one and a classical one using the simulation algorithm of Section~\ref{Calgo}. 

\begin{figure}
\begin{center}
\vspace{10pt}
\includegraphics[width=5.0cm]{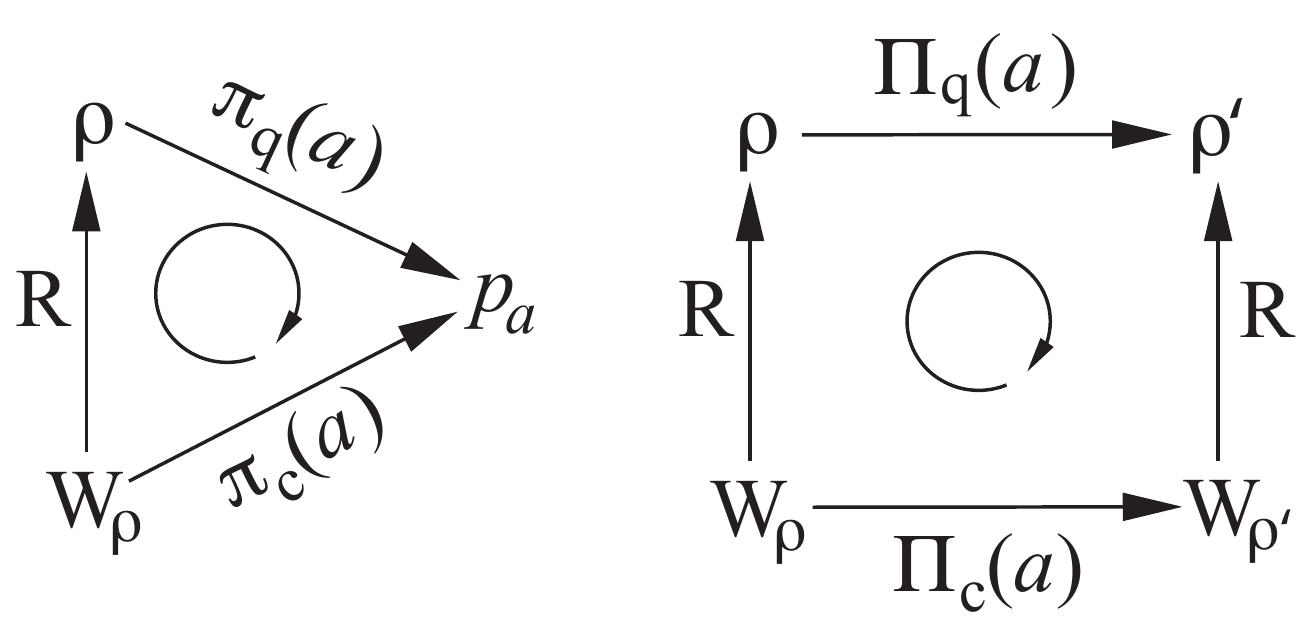}
\caption{\label{cD}Diagrams representing the quantum and the classical way of calculating the probability of measurement outcomes (left) and the post-measurement state (right).}
\end{center}
\end{figure}

Regarding the outcome probability $p_a(s_a)$ given $W_\rho$, the quantum mechanical way first obtains the corresponding quantum state $\rho$ from $W_\rho$ through Eq.~(\ref{Wigner}). This is represented by a map $R: W_\rho \mapsto \rho$. Second, from $\rho$, the outcome probability $p_a(s_a)$ is obtained via the Born rule, $p_a(s_a)=\text{Tr}(P_a(s_a)\rho)$. This is represented by a map $\pi_q(a): \rho \mapsto p_a$. The classical way uses the algorithm of Section~\ref{Calgo} to obtain $p_a$. This is represented by a map $\pi_c(a):W_\rho \mapsto p_a$.

Likewise, the quantum mechanical way of obtaining the post-measurement state $\rho'$ from $W_\rho$ proceeds by first applying the map $R$ (Eq.~(\ref{Wigner})) to obtain $\rho$, and second by obtaining $\rho'$ from $\rho$ through the Dirac projection postulate. The second step is represented by a map $\Pi_q(a)$. 

The classical way of obtaining $\rho'$ from $W_\rho$ proceeds by first using the simulation algorithm to obtain $W_{\rho'}$, and second by mapping $W_{\rho'}$ to $\rho'$ using the map $R$. The first step in this procedure is represented by the map $\Pi_c(a)$.

The classical simulation algorithm of Section~\ref{Calgo} is correct only if the quantum and the classical ways of computing $p_a(s_a)$ and $\rho'$ agree. That is, we require the diagrams in Fig.~\ref{cD} to commute.

\begin{Lemma}\label{cDL}
The diagrams of Fig.~\ref{cD} commute.
\end{Lemma}

{\em{Proof of Lemma~\ref{cDL}.}} We discuss the outcome probability and the post-measurement state separately.

{\em{Outcome probability $p_a(s_a)$.}} Then, the quantum mechanical expression for $p_{a}(s_a)$ is
\begin{equation}\label{ProbExpr}
\begin{array}{rcl}
p_a(s_a) &=& \displaystyle{\sum_{(\Omega,\gamma)\in {\cal{V}}} \!\!\!W_\rho(\Omega,\gamma)\, \text{Tr}\left( \frac{I+(-1)^{s_a}T_a}{2} A_\Omega^\gamma \right)}\\
&=& \displaystyle{\sum_{(\Omega,\gamma)\in {\cal{V}}} \!\!\!W_\rho(\Omega,\gamma) \left( \delta_{a\in\Omega}\delta_{s_a,\gamma(a)} + \frac{1}{2} \overline{\delta}_{a\in\Omega} \right).}
\end{array}
\end{equation}
The classical expression $p_a^{(c)}(s_a)$ for $p_a(s_a)$ obtained through the algorithm of Section~\ref{Calgo} is as follows. If $a\in \Omega$, then the conditional probability for the outcome $s_a$ given the state $(\Omega,\gamma)$ is $\delta_{s_a,\gamma(a)}$. If $a\not \in \Omega$ then the conditional probability for the outcome $s_a$ is 1/2. Thus,
$$p_a^{(c)}(s_a) = \sum_{(\Omega,\gamma)\in {\cal{V}}} W_\rho(\Omega,\gamma) \left( \delta_{a\in\Omega}\delta_{s_a,\gamma(a)} + \frac{1}{2} \overline{\delta}_{a\in\Omega} \right).
$$
By comparing the two expressions, we find that $p_a(s_a) = p_a^{(c)}(s_a)$ for all $a$, $s_a$, and the left diagram of Eq.~(\ref{cD}) thus commutes.

{\em{Post-measurement state $\rho'$.}} The quantum mechanical expression for the post-measurement state $\rho'$ has already been given in Eq.~(\ref{PMS}), and we now derive the corresponding expression $\rho'_{(c)}$ that follows from the classical simulation algorithm. 

We consider the joint probability $p((\Omega',\gamma')\cap s_a)$ of obtaining the outcome $s_a$ in the measurement of $T_a$ and ending up in the state $(\Omega',\gamma')$. We may invoke conditional probabilities in two ways,
\begin{widetext}
$$
p((\Omega',\gamma')\cap s_a) =\displaystyle{\sum_{(\Omega,\gamma)\in {\cal{V}}} p((\Omega',\gamma')\cap s_a|(\Omega,\gamma)) W_\rho(\Omega,\gamma)} = \displaystyle{p((\Omega',\gamma')| s_a) p_a(s_a).}
$$ 
Noting that $p((\Omega',\gamma')|s_a)=W_{\rho'}(\Omega',\gamma')$, and equating the two above expressions we find
\begin{equation}\label{Wup}
W_{\rho'}(\Omega',\gamma') = \sum_{(\Omega,\gamma)\in {\cal{V}}} \frac{W_\rho(\Omega,\gamma)}{p_a(s_a)} p((\Omega',\gamma')\cap s_a|(\Omega,\gamma)). 
\end{equation}
We now infer the conditional probabilities $p((\Omega',\gamma')\cap s_a|(\Omega,\gamma))$ from the classical simulation algorithm of Section~\ref{Calgo},
$$
p((\Omega',\gamma')\cap s_a|(\Omega,\gamma)) =\left\{\begin{array}{lr} \displaystyle{\frac{1}{2}\delta_{s_a,\gamma(a)}\delta_{\Omega',\Omega}\left(\delta_{\gamma',\gamma}+\delta_{\gamma',\gamma+[a,\cdot]} \right),}& \text{if } a\in \Omega,\vspace{1mm}\\
\displaystyle{\frac{1}{2} \delta_{\Omega',\Omega\times a} \delta_{\gamma',\gamma\times s_a}},& \text{if } a\not \in \Omega. \end{array} \right.
$$
Inserting this into Eq.~(\ref{Wup}), and using the resulting expression in Eq.~(\ref{Wigner}), i.e. applying the map $R$, we obtain
$$
\begin{array}{rcl}
\rho'_{(c)} &=& \displaystyle{\!\!\!\!\sum_{(\Omega',\gamma')\in {\cal{V}}}\sum_{(\Omega,\gamma)\in {\cal{V}}}  \frac{W_\rho(\Omega,\gamma)}{2p_a(s_a)} \left(
\delta_{a\in\Omega} \delta_{s_a,\gamma(a)}\delta_{\Omega',\Omega}\left(\delta_{\gamma',\gamma}+\delta_{\gamma',\gamma+[a,\cdot]} \right)+\overline{\delta}_{a\in\Omega} \delta_{\Omega',\Omega\times a} \delta_{\gamma',\gamma\times s_a}\right) A_{\Omega'}^{\gamma'}}\vspace{1mm}\\
&=& 
\displaystyle{\!\!\!\! \sum_{(\Omega,\gamma)\in {\cal{V}}}  \frac{W_\rho(\Omega,\gamma)}{2p_a(s_a)} \left( \delta_{a\in\Omega} \delta_{s_a,\gamma(a)} \left( A_\Omega^\gamma + A_\Omega^{\gamma+ [a,\cdot]}\right) +\overline{\delta}_{a\in\Omega} A_{\Omega\times a}^{\gamma \times s_a} \right).}
\end{array}
$$
\end{widetext}
Comparing the last expression with Eq.~(\ref{PMS}), we find that $\rho'_{(c)}=\rho'$ for all $a$, $s_a$, and the right diagram in Fig.~\ref{cD} thus commutes. $\Box$\medskip

{\em{Proof of Theorem~\ref{T1}.}} As explained in Section~\ref{QUP},  we only need to discuss sequences of Pauli measurements. For those, we show that the algorithm of Table~\ref{SimAlg} is correct, and, if the initial $W_\rho$ can be efficiently sampled from, it is also computationally efficient. (i) {\em{Correctness.}} Denote by $\rho(t)$ the state before the $t$-th measurement. With Lemma~\ref{cDL}, by induction on the right diagram in Eq.~(\ref{cD}), if $W_{\rho(1)}$ represents the initial state $\rho(1)$, then $W_{\rho(t)}$ represents $\rho(t)$ for all time steps $t=1,..,N$. Then, by the left diagram in Eq.~(\ref{cD}), the outcome probabilities $p_{a_t}(s_{a_t}|\textbf{s}_{\prec t})$, with $\textbf{s}_{\prec t}=(s_{a_1},.., s_{a_{t-1}})$ the measurement record prior to time $t$, are also correct. Thus the joint outcome probability  sampled from
$$
p_{a_1,..,a_n}(s_{a_1},..,s_{a_N}) = \prod_{t=1}^N p_{a_t}(s_{a_t}|\textbf{s}_{\prec t}),
$$
is also correct.

(ii)  {\em{Efficiency.}} We recall that all cnc sets~$\Omega$ are unions of~$O(n)$ isotropic spaces $\Omega_i$  (Theorem~\ref{nms}). Further, each~$\Omega_i$ defines a stabilizer group
\begin{equation}\label{stabs}
T_{\Omega_i}^\gamma:=\{T_a^\gamma:=(-1)^{\gamma(a)}T_a,a\in\Omega_i\}.
\end{equation}
This allows us to describe~$(\Omega,\gamma)\in\mathcal{V}$ using polynomial memory by storing~$O(n)$ stabilizer tables of size~$O(n^2)$~\cite{Goma,Gottesman99Heisenberg}. Indeed, by Defs.~\ref{Def_Cl}-\ref{Def_NC} and Lemma~\ref{L4},~$T_{\Omega_i}^\gamma$ is a closed commutative group. Furthermore, with Def.~\ref{Def_NC}, it holds that $T_a^\gamma T_b^\gamma=T_{a+b}^\gamma, \forall a,b \in \Omega_i$. This implies the existence of a non-trivial stabilized subspace: $P_{\Omega_i}^\gamma:=\sum_{a\in \Omega_i} T_{a}^\gamma/|\Omega_i|$ is a common +1-eigenprojector of every~$T_a\in T_{\Omega_i}^\gamma$ as~$ T_a^\gamma P_{\Omega_i}^\gamma  = \sum_{b\in \Omega_i} \frac{T_{a+b}^\gamma}{|\Omega_i|}=  \sum_{b'\in \Omega_i} \frac{T_{b'}^\gamma}{|\Omega_i|} = P_{\Omega_i}^\gamma, \forall T_{\Omega_i}^\gamma$, which also implies ${P_{\Omega_i}^{\gamma}}^2 =  P_{\Omega_i}^\gamma$.

We now note that the update rules in algorithm~\ref{SimAlg}, namely (i) checking whether $a \in \Omega$, (ii) evaluating $\gamma$ on $a\in \Omega$, (iii) updating $\gamma \longrightarrow \gamma + [a,\cdot]$, (iv) $\Omega \longrightarrow \Omega \times a$ and (v) $\gamma \longrightarrow \gamma\times s_a$, implement  tasks that admit efficient classical algorithms in the stabilizer formalism~\cite{Goma,Gottesman99Heisenberg}. 
\emph{Rules (i) and (ii):} To test~$a\in\Omega$, we check whether~$a\in \Omega_i, i=1,\ldots,O(n)$. If $a\in\Omega_j$ for some value of~$j$, then $\gamma(a)$ is computed as the bit determining the phase of the stabilizer operator~$T_a^\gamma\in T_{\Omega_i}^\gamma$. Both tasks can be solved  classically efficiently via Gaussian elimination given the stabilizer table data~\cite{Goma,Gottesman99Heisenberg}. \emph{Rule (iii):} $\gamma$ is updated to~$\gamma'=\gamma+[a,\cdot]$ by (classically efficiently) evaluating~$\gamma(\cdot)+[a,\cdot]$ on the generators of every~$\Omega_i$. \emph{Rules (iv) and (v):} For all $j$, $T_{\Omega_j \times a}^{\gamma|_{\Omega_j} \times {s_a}}$ is the stabilizer group resulting from the measurement of $T_a$ with outcome $s_a$ on a state with stabilizer group $T_{\Omega_j}^{\gamma|_{\Omega_j}}$. This update can be efficiently performed using the standard measurement update-rule of Ref.~\cite{Goma,Gottesman99Heisenberg} to every stabilizer table in the description of ~$(\Omega,\gamma)$. Thus, all steps of the algorithm run in polynomial time. $\Box$

\section{The case of $W_\rho< 0$}\label{CEsimNeg}

As we have established in the previous sections, $W_\rho< 0$ is a precondition for quantum speedup. When the initial state is represented by a quasiprobability rather than a true probability function, a standard problem of interest is estimating outcome probabilities for sequences of measurements. An established method for probability estimation is \cite{Pashayan}, utilizing the Hoeffding bound. Note that probability estimation is a different problem than weak simulation \cite{VdN1}, and is not efficiently adaptive.

\subsection{Robustness}

In close analogy to the ``robustness of magic'' \cite{RoM} $\mathfrak{R}_S$ (the subscript $S$ is for ``stabilizer''), we define a phase space robustness $\mathfrak{R}$, through
\begin{equation}\label{Def_Rob}
\mathfrak{R}(\rho):=\min_{W|\, \langle{\cal{A}},W\rangle= \rho}\|W\|_1,
\end{equation}
with $\langle{\cal{A}},W\rangle:=\sum_{\alpha \in {\cal{V}}} W_\alpha A_\alpha$. 

Since the definitions of the robustness $\mathfrak{R}$ and of the robustness of magic $\mathfrak{R}_S$ \cite{RoM} are so similar, one may wonder if there is a relation between them. This is indeed the case; namely, we have the following result.

\begin{Lemma}\label{relR}
For all quantum states $\rho$, of any number $n$ of qubits, the phase space robustness $\mathfrak{R}(\rho)$ and the robustness of magic $\mathfrak{R}_S(\rho)$ are related via
\begin{equation}\label{RobRel}
\mathfrak{R}(\rho) \leq \mathfrak{R}_S(\rho) \leq (4n+1)\, \mathfrak{R}(\rho).
\end{equation}
\end{Lemma}
Thus, the phase space robustness $\mathfrak{R}$ is never larger than the robustness of magic, but can only be moderately smaller. The proof of Lemma~\ref{relR} is given in Appendix~\ref{ProofRelR}.

\subsection{Hardness of classical simulation}

The Hoeffding bound says that the number $N$ of samples required to estimate the output probability distribution up to an error $\epsilon$ scales as
$N \sim {\cal{M}}^2/\epsilon^2$, where ${\cal{M}}$ is a measure of the negativity contained in the quantum process. 
In our case, the operations are positivity-preserving, and all negativity comes from the initial state. The algorithm of Pashayan et al. \cite{Pashayan}, when applied to our setting, says that the number $N$ of samples required to estimate the output probability scales as
$$
N \sim \frac{\mathfrak{R}(\rho_\text{init})^2}{\epsilon^2}.
$$
Thus, the robustness $\mathfrak{R}(\rho_\text{init})$ of the initial state $\rho_\text{init}$ is the critical parameter determining the classical hardness of probability estimation. 

The same relation, with the robustness $\mathfrak{R}$ replaced by the robustness of magic $\mathfrak{R}_S$ holds for the classical simulation based on quasiprobability distributions over stabilizer states \cite{RoM}. Lemma~\ref{relR} above is therefore of interest for relating the operational costs of the two simulation methods.\medskip

Classical simulation also requires a quasiprobability function $W_{\mu^{\otimes n}}$ for $n$ copies of the magic state $\mu$. Since the $n$-qubit phase space is large, the numerical optimization to obtain the least-negative expansion $W_{\mu^{\otimes n}}$ is computationally costly. However, we can apply a similar splitting into smaller blocks of magic states as in the stabilizer case \cite{RoM}. The computational cost for providing the expansion is then a function of block size rather than total number of copies $n$. The 1-norm of the resulting expansion is smaller than of the stabilizer expansion, by a factor that is constant in $n$. Details are given in Appendix~\ref{LTpr}.

\subsection{Elements of a resource theory based on $W$}\label{PhaSpaStruct}

It is illuminating to discuss QCM within the framework of resource theories. Every resource theory has three main operational components \cite{BG15}, (i) the resource(s), (ii) the non-resources, or free states, (iii) the free operations. 

In the physical setting of our interest, the resources are quantum states  which cannot be positively represented by $W$ (cf. Theorem~\ref{T1}). The free operations are Clifford unitaries and Pauli measurements. The free states are those that can be created from the free operations  from a completely mixed state, i.e., all mixtures of stabilizer states.

We observe that there is a third class of states which are neither resources nor free, namely the positively representable states which are not mixtures of stabilizer states. Such states are called (iv) bound magic states. We have seen an example of them in Section~\ref{Exa}, the general 1-qubit states tensored with a stabilizer state on arbitrarily many qubits.

The reason for calling those states ``bound magic'' is that they cannot be distilled into computationally useful ones by free operations. In our setting, by Theorem~\ref{CPM}, positive representability is an invariant under the free operations. Hence, bound states can only be converted into other bound states or into free states by the free operations, but never into a resource.

The question of inter-convertibility may more generally be asked for resource states. To facilitate this discussion, one may identify monotones, i.e., real-valued functions on the state space that never increase under the free operations. The main result of this section is that the robustness $\mathfrak{R}$, defined in Eq.~(\ref{Def_Rob}) and already known to measure hardness of classical simulation by sampling, is a monotone. 

\begin{Theorem}\label{Mono}
The robustness $\mathfrak{R}$ is a monotone under all Clifford unitaries and Pauli measurements.
\end{Theorem}
As part of the proof of Theorem~\ref{Mono}, we now discuss an important structural property of the quasiprobability function $W$, namely its covariance under Clifford unitaries.
Be $\text{Cl}_n$ the $n$-qubit Clifford group. It acts on the $n$-qubit Pauli operators via
$$
h(T_a):= h T_a h^\dagger = (-1)^{\Phi_h(a)}T_{ha},\;\; \forall h \in \text{Cl}_n.
$$  
This relation simultaneously defines the phase function $\Phi$ and the action of $\text{Cl}_n$ on $E$.
It implies an action of the Clifford group on the phase point operators $A_\Omega^\gamma$, which in turn induces an action on the sets $\Omega$ and the functions $\gamma$, via
$$
h(A_\Omega^\gamma) = \frac{1}{2^n} \sum_{a\in \Omega} (-1)^{\gamma(a)}h(T_a) = \frac{1}{2^n}\sum_{b \in \Omega'}(-1)^{\gamma'(b)}T_b.
$$
Therein, the set $\Omega'$ is defined as $\Omega' := \{ha,\; a\in \Omega\}$, and the function $\gamma':\Omega'\longrightarrow \mathbb{Z}_2$ is given by
$$
\gamma'(ha):= \gamma(a)+\Phi_h(a),\;\forall a\in \Omega.
$$
Henceforth we denote $\Omega'$ as $h\cdot \Omega$ and $\gamma'$ as $h\cdot \gamma$, to emphasize the dependence on $h \in \text{Cl}_n$. 

For use in the proof below we quote Lemma 3 from \cite{Coho} which says that, for any face  $(a,b)\in \Omega \times \Omega$, 
$$
\Phi_h(\partial (a,b)) = \beta(ha,hb) +\beta(a,b) \mod 2.
$$
We then have the following result.
\begin{Lemma}\label{Covar}
${\cal{V}}$ is mapped to itself under $\text{Cl}_n$, and the quasiprobability function $W$ transforms covariantly. That is, if the state $\rho$ can be described by $W_\rho$ through Eq.~(\ref{Wigner}), then for any $h\in \text{Cl}_n$ the state $h\rho h^\dagger$ can be described by a quasiprobability function $W_{h\rho h^\dagger}$ defined by
$$
W_{h \rho h^\dagger}(\Omega,\gamma):= W_\rho(h^{-1}\cdot \Omega,h^{-1}\cdot \gamma).
$$ 
\end{Lemma}

{\em{Remark 3:}} We say ``the state $\rho$ can be described by $W_\rho$" rather than ``is described'' because $W_\rho$ is not unique. \medskip

{\em{Proof of Lemma~\ref{Covar}.}} First, we show that the phase space ${\cal{V}}$ is closed under the action of $\text{Cl}_n$, i.e., if $(\Omega,\gamma) \in {\cal{V}}$ then $(\Omega',\gamma') \in {\cal{V}}$. The four items in Definition~\ref{PhaSpa} need to be checked. (i) {\em{Closedness under inference.}} Assume that $c,d\in \Omega'$ and $[c,d]=0$. Then there exist $a,b \in \Omega$ such that $c=ha$, $d=hb$ and $[a,b]=0$. Then, $c+d = ha+hb = h(a+b) \in \Omega'$, since $a+b\in \Omega$ by the assumption of closedness. Hence $\Omega'$ is closed under inference.  

(iii) {\em{$\gamma'$ satisfies Eq.~(\ref{dgb}).}} With the definition of $\gamma'$ we have (all addition mod 2)
$$
\begin{array}{rcl}
d\gamma'(ha,hb) &=& d\gamma(a,b) + \Phi_h(\partial(a,b))\\
&=&  d\gamma(a,b) +  \beta(ha,hb) +\beta(a,b)\\
&=& \beta(ha,hb).
\end{array}
$$
Therein, in the second line we have used Eq.~(\ref{dgb}). Thus, $\gamma'$ satisfies Eq.~(\ref{dgb}) on its domain.

(ii) {\em{$\Omega'$ is non-contextual.}} With $\gamma'$ we have just proved the existence of a function on $\Omega'$ that satisfies Eq.~(\ref{dgb}).

(iv) {\em{$\gamma'$ satisfies Eq.~(\ref{NormCond}).}} Since $\gamma$ satisfies Eq.~(\ref{NormCond}), it follows $I=h(I)=h\left((-1)^{\gamma(0)}T_0\right) = (-1)^{\gamma(0)+\Phi_h(0)}T_0=(-1)^{\gamma'(0)}T_0$. Eq.~(\ref{NormCond}) is thus satisfied for $\gamma'$.

Hence, if $(\Omega,\gamma)\in {\cal{V}}$ then $(\Omega',\gamma')\in {\cal{V}}$, as claimed.\smallskip 

Next we turn to the covariance of $W$ under $\text{Cl}_n$. We  have
$$
\begin{array}{rcl}
h\rho h^\dagger &=& \sum_{(\Omega,\gamma) \in {\cal{V}}} W_\rho(\Omega,\gamma) h(A_\Omega^\gamma)\\
&=& \sum_{(\Omega,\gamma) \in {\cal{V}}} W_\rho(\Omega,\gamma) A_{h\cdot \Omega}^{h \cdot \gamma}\\
&=& \sum_{(\Omega,\gamma) \in {\cal{V}}} W_\rho(h^{-1} \cdot\Omega,h^{-1}\cdot \gamma) A_{\Omega}^{\gamma}.
\end{array}
$$
Comparing the last expression with the expansion Eq.~(\ref{Wigner}) for $h\rho h^\dagger$,  we find that for all $h \in \text{Cl}_n$, the quasiprobability distribution $W_{h\rho h^\dagger}$ defined by 
\begin{equation}\label{CovCond}
W_{h\rho h^\dagger}(\Omega,\gamma) = W_\rho(h^{-1}\cdot \Omega,h^{-1}\cdot \gamma) \end{equation}
describes the state $h\rho h^\dagger$. This is the covariance condition. $\Box$\medskip

\noindent
We are now ready to prove the monotonicity of $\mathfrak{R}$, as stated in Theorem~\ref{Mono}.\smallskip

{\em{Proof of Theorem~\ref{Mono}.}} (a) Clifford unitaries. With Lemma~\ref{Covar}, we have that for any $n$-qubit Clifford gate $h$ applied to any $n$-qubit state $\rho$, the quasiprobability distribution $W_{h\rho h^\dagger}$ can be related to $W_\rho$ via the covariance condition Eq.~(\ref{CovCond}). Since $W$ is non-unique, there may a priori be a representation $W'_{h\rho h^\dagger}$ with smaller 1-norm, and thus it holds that
\begin{equation}
\mathfrak{R}(h\rho h^\dagger) \leq \mathfrak{R}(\rho),\;\; \forall \rho,\;\forall h \in \text{Cl}_n.
\end{equation}

(b) Pauli measurements. We consider the measurement of a Pauli observable $T_a$ on a quantum state $\rho$. Denote by $\rho_{a,s_a}$ the normalized post-measurement states for the outcomes $s_a=0,1$, respectively. We have to show that, for all $n$, for all $a \in \mathbb{Z}_2^n \times \mathbb{Z}_2^n$ and all $n$-qubit states $\rho$ it holds that
\begin{equation}\label{MonoMeas}
p_a(0) \,\mathfrak{R}(\rho_{a,0}) + p_a(1)\, \mathfrak{R}(\rho_{a,1}) \leq \mathfrak{R}(\rho).
\end{equation}
With Eq.~(\ref{W_elements}), we can write $p_a(0)\,W_{\rho_{a,0}}= W_+ + \overline{W}_+$, and $p_a(1)\,W_{\rho_{a,1}}= W_- + \overline{W}_-$, where
\begin{widetext}
\begin{equation}\label{W_blocks}
\begin{array}{rcl}
W_+(\Omega',\gamma')&:=&  \sum_{(\Omega,\gamma) \in {\cal{V}}}\frac{W_\rho(\Omega,\gamma)}{2} \delta_{a\in \Omega} \delta_{\gamma(a),0}\left(\delta_{(\Omega',\gamma'),(\Omega,\gamma)}+ \delta_{(\Omega',\gamma'),(\Omega,\gamma+[a,\cdot])} \right),\vspace{2mm}\\

W_-(\Omega',\gamma')&:=&  \sum_{(\Omega,\gamma) \in {\cal{V}}}\frac{W_\rho(\Omega,\gamma)}{2} \delta_{a\in \Omega} \delta_{\gamma(a),1}\left(\delta_{(\Omega',\gamma'),(\Omega,\gamma)}+ \delta_{(\Omega',\gamma'),(\Omega\,\gamma+[a,\cdot])} \right),\vspace{2mm}\\

\overline{W}_+(\Omega',\gamma')&:=& \sum_{(\Omega,\gamma) \in {\cal{V}}}\frac{W_\rho(\Omega,\gamma)}{2} \overline{\delta}_{a\in \Omega}\delta_{(\Omega',\gamma'),(\Omega \times a,\gamma \times (s_a =0))},\vspace{2mm}\\
\overline{W}_-(\Omega',\gamma')&:=& \sum_{(\Omega,\gamma) \in {\cal{V}}}\frac{W_\rho(\Omega,\gamma)}{2} \overline{\delta}_{a\in \Omega}\delta_{(\Omega',\gamma'),(\Omega \times a,\gamma \times (s_a =1))}.
\end{array}
\end{equation}
\end{widetext}
From now on, denote by $W_\rho$ the optimal representation for $\rho$ w.r.t. 1-norm, i.e., $\mathfrak{R}(\rho)=\left\|W_\rho\right\|_1$. With the triangle inequality, and the fact that the functions $W_{\rho_{a,s_a}}$ induced from the optimal $W_\rho$ through Eq.~(\ref{W_blocks}) need not be optimal for the states $\rho_{a,s_a}$ w.r.t. their 1-norm, it holds that
$p_{a,0}\, \mathfrak{R}(\rho_{a,0}) \leq \left\|W_+\right\|_1 + \left\|\overline{W}_+\right\|_1$, and $p_{a,1}\, \mathfrak{R}(\rho_{a,1}) \leq \left\|W_-\right\|_1 + \left\|\overline{W}_-\right\|_1$, hence
\begin{equation}\label{NormRel}
\begin{array}{rcl}
p_{a,0}\, \mathfrak{R}(\rho_{a,0}) + p_{a,1}\, \mathfrak{R}(\rho_{a,1}) &\leq& \left\|W_+\right\|_1 +   \left\|W_-\right\|_1+\\
&&  \left\|\overline{W}_+\right\|_1 + \left\|\overline{W}_-\right\|_1.
\end{array}
\end{equation}
With Eq.~(\ref{W_blocks}) we find that
$$
\begin{array}{l}
\left\|W_+\right\|_1 + \left\| W_-\right\|_1 =\vspace{1mm}\\
= \sum_{(\Omega',\gamma')\in {\cal{V}}} \frac{\delta_{a\in \Omega'}}{2} |W_\rho(\Omega',\gamma')+W_\rho(\Omega',\gamma'+[a,\cdot])|\vspace{1mm}\\
\leq  \sum_{(\Omega',\gamma')\in {\cal{V}}} \delta_{a\in \Omega'}|W_\rho(\Omega',\gamma')|,
\end{array}
$$
where in the second line we used the triangle inequality again. Furthermore, performing the summation over all $(\Omega',\gamma')\in {\cal{V}}$ first, we obtain
$$
\left\|\overline{W}_+\right\|_1 = \left\| \overline{W}_-\right\|_1 \leq \sum_{(\Omega,\gamma)\in {\cal{V}}} \frac{\overline{\delta}_{a\in \Omega}}{2} |W_\rho(\Omega,\gamma)|.
$$
Inserting the last two relations into Ineq.~(\ref{NormRel}), we arrive at
$$
p_{a,0}\, \mathfrak{R}(\rho_{a,0}) + p_{a,1}\, \mathfrak{R}(\rho_{a,1}) \leq \left\| W_\rho \right\|_1.
$$
Since $\mathfrak{R}(\rho)=\left\|W_\rho\right\|_1$ by assumption, Eq.~(\ref{MonoMeas}) follows. $\Box$ 

\subsection{Numerical results}

\begin{figure}
	\begin{center}
		\includegraphics[width=7.8cm]{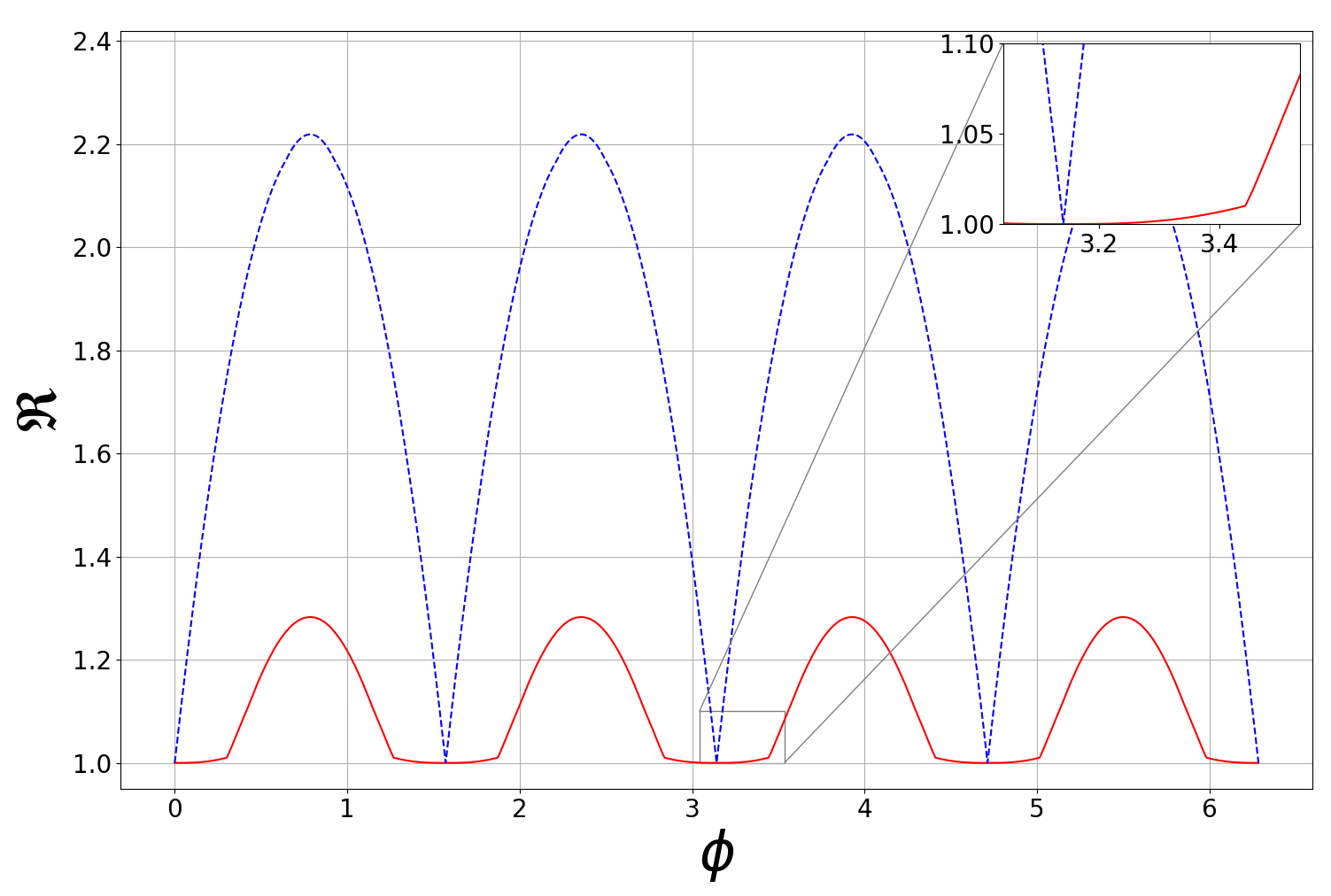} 
		\caption{\label{Robust} Robustness $\mathfrak{R}$ (solid line) and robustness of magic $\mathfrak{R}_S$ (dotted line) for the state $|H(\phi)\rangle^{\otimes 3}$, cf. Eq.~(\ref{Tphi}), as a function of $\phi$. Highlighted is the region near a stabilizer state, at $\phi= \pi$.} 
	\end{center}
\end{figure}

In Table~\ref{tab:robustness} and Fig.~\ref{Robust} we present numerical values \footnote{Our calculations use  the software packages CVXPY \cite{cvxpy} and GUROBI \cite{gurobi}.} for the robustness of various magic states, and compare them to robustness of magic as defined by Howard and Campbell \cite{RoM}.  Table \ref{tab:robustness} summarizes the robustness comparisons for the common magic states, as well as the maximal-robustness Hoggar state \cite{RoM}. In Fig.~\ref{Robust} we plot the robustness against the stabilizer state robustness for three qubits, as a function of rotation angle. Note the wide and almost flat---though not perfectly flat---plateaus of robustness $\mathfrak{R}$ in the vicinity of stabilizer states.

\begin{table}
		\begin{center}
		\begin{tabular}{|l|l|l|}\hline
			state	& $\mathfrak{R}$ & $\mathfrak{R}_S$ \\ \hline
			$\ket{H}^{\otimes 2}$	& 1.0  &  1.7472   \\ \hline
			$\ket{T}^{\otimes 2}$	& 1.0 & 2.23205    \\ \hline
			$\ket{H}^{\otimes 3}$	& 1.283  & 2.2189  \\\hline
			$\ket{T}^{\otimes 3}$	& 1.385 & 3.09807  \\ \hline
			$\ket{\text{Hoggar}}$ & 1.80 & 3.8000   \\ \hline
		\end{tabular}
		\caption{\label{tab:robustness}Robustness values of selected magic states. For robustness of magic ($\mathfrak{R}_S$), also see \cite{Heinrich}.}
	\end{center}
\end{table}

\subsection{Curious resurgence of $4^n$-dimensional phase space}

Numerical calculations of robustness for various quantum states revealed an unexpected feature. Namely, the optimal quasiprobability distribution $W_\rho$ w.r.t. Eq.~(\ref{Def_Rob}) for a given $n$-qubit state $\rho$ always was non-zero only on $4^n$  phase space points, or fewer. $4^n$ is only a tiny fraction of the whole phase space ${\cal{V}}$, and furthermore the naive expectation if one were completely oblivious of the differences between even and odd $d$. However, the support of the optimal $W(\rho)$ depends on the state $\rho$. 
We can now explain the initially puzzling upper bound on the size of the support, $4^n$.\smallskip

The robustness $\mathfrak{R}$ of a state $\rho$ defined in Eq.~(\ref{Def_Rob}) is the solution to the convex optimization problem
\begin{equation}\label{convRob}
\min\limits_{\textbf{q}}\left\{||\textbf{q}||_1:\;M\textbf{q}=\textbf{b}\right\},
\end{equation}
where $M_{i,j}=\text{Tr}(A_{\alpha_j}P_i)$, $b_i=\text{Tr}(\rho P_i)$, $\{\alpha_j:1\le j\le|\mathcal{V}|\}$ is an enumeration of the phase points and $P_i$ are the $n$-qubit Pauli operators.
For each variable $q_j$ in Eq.~({\ref{convRob}}), define two new variables $q_j^+\coloneqq\max(0, q_j)$ and $q_j^-\coloneqq\max(0, -q_j)$.  Then the convex optimization problem of Eq.~(\ref{convRob}) is equivalent to the standard form linear program
\begin{equation}\label{LPRob}
\min\limits_{\textbf{q}}\left\{\sum\limits_{j}q_j^++q_j^-:\;\tilde{M}\tilde{\textbf{q}}=\textbf{b},\;\tilde{\textbf{q}}\ge0\right\},
\end{equation}
where $\tilde{M}=\begin{bmatrix}M & -M\end{bmatrix}$ and $\tilde{\textbf{q}}=\begin{bmatrix}(\textbf{q}^+)^T & (\textbf{q}^-)^T\end{bmatrix}^T$.  This doubles the number of variables but does not change the number of equality constraints.  Since we know this problem is feasible (any physical state can be written as an affine combination of phase point operators) and bounded (no physical state can have robustness less than 1), by the fundamental theorem of linear programming, for any physical state, Eq.~({\ref{LPRob}}) has a solution at a vertex of the feasible polytope~{\cite{opt}}.

Since Eq.~({\ref{LPRob}}) has an equality constraint for each $n$-qubit Pauli operator (including the identity), this means any state $\rho$ has a robustness-minimizing expansion in phase point operators with no more than $4^n$ non-zero coefficients. 

\section{Discussion}\label{Disc}

\subsection{Stratonovich-Weyl correspondence}\label{SW}

In the field of quantum optics, an important set of
criteria for a proper quasiprobability distribution over a phase space
is given by the Stratonovich-Weyl (SW) correspondence. Denote by $F_A^{(s)}: X \longrightarrow \mathbb{R}$ the quasiprobability distribution corresponding to the operator $A$, with $X$ the phase space and $s$ a real parameter in the interval $[-1,1]$. In the standard formalism for infinite-dimensional Hilbert spaces, $s=-1,0,1$ correspond to the Glauber-Sudarshan $P$, Wigner, and Husimi $Q$ function, respectively. The SW correspondence is the following set of criteria on the $F_A^{(s)}$ \cite{Strato}; also see  \cite{Brif}, 
\begin{enumerate}
\item[(0)]{Linearity: $A \longrightarrow F_A^{(s)}$ is a one-to-one linear map.}
\item[(1)]{Reality: $$F_{A^\dagger}^{(s)}(u) = \left(F_A^{(s)}(u)\right)^*, \forall u \in X.$$}
\item[(2)]{Standardization: $$\int_X d\mu(u) F_A^{(s)}(u) = \text{Tr}\, A.$$}
\item[(3)]{Covariance: $$F^{(s)}_{g\cdot A}(u) = F^{(s)}_A(g^{-1} u), \; g\in G,$$ with $G$ the dynamical symmetry group.}
\item[(4)]{Traciality: $$\int_X d\mu(u) F_A^{(s)}(u)F_B^{(-s)}(u) = \text{Tr}\, AB.$$}
\end{enumerate}
We now investigate to which extent these SW criteria apply to the present quasiprobability function $W$. There are two deviations. First, the present quasiprobability function $W$ does not come with a parameter $s$; there is only a single function $W$. This will affect the formulation of traciality. Second, the present mapping $A \longrightarrow W_A$ is one-to-many, as we have noted in Section~\ref{PhaseSp}. The mapping is nonetheless linear, $A+B$ can be represented as $W_A+W_B$.

The remaining SW conditions do apply. (1) Reality of $W$ follows directly from the definition Eq.~(\ref{Wigner}), since all $A_\Omega^\gamma$ are Hermitian. (2) Standardization: The definition Eq.~(\ref{PPO2}) and property Eq.~(\ref{NormCond}) of the phase point operators imply $\text{Tr}\, A_\Omega^\gamma=1$, for all $\Omega$ and $\gamma$; and standardization then follows from Eq.~(\ref{Wigner}). 

(3) Covariance holds for the entire Clifford group, as stated in Lemma~\ref{Covar}. In fact, insisting on Clifford covariance leads to the non-uniqueness of $W$. Namely, an over-complete set of phase point operators is necessary to achieve Clifford covariance \cite{Zhu}.

(4) Traciality. In the absence of a continuously varying parameter $s$, we define a dual quasiprobability function $\tilde{W}$ in addition to $W$, to stand in for $F^{(-s)}$. For all Pauli operators $T_a$ we have 
$$
\tilde{W}_{T_a}(\Omega,\gamma):=\left\{ \begin{array}{ll}  (-1)^{\gamma(a)}, & a\in \Omega\\ 0,& a \not\in \Omega \end{array}\right. .
$$
Since the $n$-qubit Pauli operators form an operator basis, $\tilde{W}$ can be extended to all $n$ qubit operators by linearity. With Eq.~(\ref{ProbExpr}) we then have
$$
\text{Tr}\, A\rho = \sum_{(\Omega,\gamma)\in {\cal{V}}} \tilde{W}_A(\Omega, \gamma) W_\rho(\Omega,\gamma). 
$$
We thus satisfy the SW criteria (1) - (4).

To conclude, we reiterate that for the present purpose of classically simulating QCM, a crucial property of $W$ is positivity preservation under Pauli measurement. This property has no counterpart in the Stratonovich-Weyl correspondence.

\subsection{Probabilistic hidden variable model}\label{pHVM}

In the case of odd $d$ \cite{NegWi}, there is a third equivalent indicator of classicality, next to positivity of the initial Wigner function and the efficiency of classical simulation of QCM by sampling. Namely, a positive Wigner function is equivalent to a non-contextual hidden variable model (HVM) with deterministic value assignments \cite{Howard}. This triple coincidence cannot be replicated in $d=2$, because, for $n\geq 2$ all quantum states---even the completely mixed state---are contextual \cite{Howard}.  

One interpretation of this situation is that contextuality, i.e., the unviability of non-contextual HVMs, is not sufficiently tight a criterion to reveal genuine quantumness. A more stringent marker is required, which (i) classifies the present HVM as classical, and (ii) for QCM in odd $d$ reduces to contextuality. At present, we have no suggestion for this more restrictive notion of quantumness. However, we point to a hidden variable model that is illustrative of the shifted quantum-to-classical boundary in the multi-qubit case, and we propose it for further study.

Namely, when positive, the quasiprobability distribution $W$ can be considered an HVM. While classified as contextual by the common definitions, it shares many features with non-contextual HVMs. 

This HVM consists of a triple~$(\Lambda, \{h^\lambda\},p_\lambda)$ where~$\Lambda = {\cal{V}}$ or~${\cal{V}}_M$,~$h^\lambda$ is a compatible family of distributions on the set of outcomes on contexts  and~$p_\lambda$ is a probability distribution on the set~$\Lambda$ of hidden variables. For each~$\alpha=(\Omega,\gamma)$ we define~$h^\alpha$   by
\begin{equation}\label{eq:HVM-def}
 h^\alpha_I(s) =
\text{Tr}(P_s A_\Omega^\gamma).
\end{equation}
Therein,~$I$ is any isotropic subspace,~$s:I\to \mathbb{Z}_2$ is a function, and~$P_s$ is the projector corresponding to the outcome.  Note that~$P_s=0$ if~$ds\neq \beta$.
	
It is useful to state the probability distributions~$h_I^\alpha(\cdot)$ in their explicit form. Let~$P_s$ denote the projector corresponding to the non-contextual value assignment~$s:I\to \mathbb{Z}_2$. Then we have
\begin{equation}\label{ValAss}
	h^{(\Omega,\gamma)}_I(s) = \frac{|I\cap \Omega|}{|I|} \delta_{s|_{I\cap \Omega},\gamma|_{I\cap \Omega}}.
\end{equation}
From Eq.~(\ref{ValAss}) we see that the value assignments in our HVM are generally probabilistic; only in the special case of~$I\subset \Omega$ they become deterministic.  Further,  the~$\{h_I^\alpha\}$ form compatible families, 
$$
h^\alpha_I|_{I\cap I'} = h^\alpha_{I'}|_{I'\cap I},\;\forall I,I',\;\forall \alpha \in {\cal{V}}.
$$
When applicable, this HVM reproduces the predictions of quantum mechanics (cf. Theorem~\ref{T1}) for measurements of Pauli observables, in single contexts or arbitrary measurement sequences.

We argue that the HVM of Eq.~(\ref{eq:HVM-def}) is classical. It is an HVM with partial value assignments, with deterministic values for some observables and random values for  others. The only resource this HVM uses beyond those required by non-contextual HVMs with deterministic value assignments is that of classical uniform randomness (in the evaluation of value assignments). Such use of randomness should not render the present HVM genuinely quantum.

And yet, the HVM of Eq.~(\ref{eq:HVM-def}) is (a) contextual in the sense of Abramsky and Brandenburger \cite{AB11}, (b) preparation and transformation contextual, as well as measurement-non-contextual, in the sense of Spekkens \cite{Spekkens}, and (c) contextual for sequences of transformations in the recently-introduced sense of Mansfield and Kashefi \cite{Mansfield18}. 
\smallskip

To summarize, we have described a hidden variable model corresponding to positive quasiprobabilities $W$. By this correspondence, the HVM is considered classical from the quantum optics perspective. It is also classical from the computational perspective, as it leads to efficient classical simulation of QCM (for applicable magic states). And yet this HVM is contextual, per the definitions commonly applied. As such, the present HVM may serve as a reference point for a refined foundational notion of quantumness that goes beyond contextuality.

\section{Conclusion}\label{Concl}

We have introduced a  quasiprobability distribution $W$ over generalized phase space, which is defined for any number $n$ of qudits with any number $d$ of levels. For multi-qudit systems with odd local dimension $d$, $W$ reduces to the familiar Wigner function for finite-dimensional systems defined by Gross \cite{Gross}. For even $d$, the phase space is enlarged and $W$ becomes non-unique. Importantly, also for $d=2$ (the multi-qubit case), $W$ has the property that a positive quasiprobability function remains positive under all Pauli measurements.  This property is crucial for classical simulation algorithms of quantum computation with magic states (QCM) by sampling.

Once this fundamental property is established, it is natural investigate the efficiency (or non-efficiency) of classical simulation in the various regimes, and resource theories characterizing QCM. Here we have treated the canonical questions that arise in this context:  we have devised an efficient classical simulation of QCM for $W\geq 0$, and clarified the relation to the qubit stabilizer formalism. Namely, the present method for efficient classical simulation of QCM strictly contains the stabilizer method. It applies to all mixtures of stabilizer states, but in addition to certain states outside the stabilizer polytope. We have further characterized the hardness of classical simulation for $W<0$ in terms of a robustness measure, and established this robustness is a monotone under the free operations of QCM. 

In summary, we arrive at a resource perspective of QCM on qubits that closely resembles the corresponding picture for odd dimension $d$. However, there are two deviations. First, the phase space on which the quasiprobability function $W$ is defined has a far more intricate structure for $d=2$ than for odd $d$. Second, for $d=2$ the hidden variable model (HVM) induced by any non-negative quasiprobability function $W_\rho$ is contextual, as a consequence of Mermin's square. 

The latter observation leads to a puzzle. The HVM induced for positively representable states $\rho$ is classified as ``classical'' from the perspectives of quantum optics ($W_\rho\geq 0$) and computer science (classical simulation is efficient), but it is classified as ``quantum'' from the perspective of contextuality. 

In this regard, we have argued (also see \cite{Howard}) that in multi-qubit QCM, contextuality is not suitable as an indicator of genuine quantumness.  We have proposed the notion of ``HVM with partial non-contextual value assignments'' in which is classicality and contextuality coexist.\smallskip 

{\em{Acknowledgments.}} We thank Piers Lillystone (JBV, CO, RR, ET) and Shane Mansfield (JBV) for discussion. This work is funded by NSERC (CO, RR, ET, MZ), Cifar (RR) and ERC (JBV).

\appendix
\section{Proof of Lemma~\ref{MaxOm}}\label{MaxOmProof}

Recall that, unlike most material in this paper, Lemma~\ref{MaxOm} holds for all local dimensions $d$.\medskip

{\em{Proof of Lemma~\ref{MaxOm}.}} Consider two sets, $\Omega,\tilde{\Omega} \in {\cal{V}}$, such that $\Omega \subset \tilde{\Omega}$, and the phase point operator
$A_\Omega^\gamma$ according to Eq.~(\ref{PPO2}). Furthermore, denote by $\tilde{\Gamma}$ the set of value assignments $\tilde{\gamma}:\tilde{\Omega}\longrightarrow \mathbb{Z}_d$ that satisfy the constraint 
$$
\tilde{\gamma}|_\Omega = \gamma.
$$
Then, $\tilde{\Gamma}$ is the coset of a vector space $U$.  This is the first fact we prove. Write $\tilde{\gamma} = \tilde{\gamma}_0 + \eta$, where $\tilde{\gamma}_0\in \tilde{\Gamma}$ is some reference function, and the functions $\eta\in U$ all satisfy
\begin{subequations}\label{etaProps}
	\begin{align}\label{EP1}
	d \eta &= 0,\\
	\label{EP2}
	\eta |_\Omega &= 0.
	\end{align}
\end{subequations}
The condition of Eq.~(\ref{EP1}) need only be satisfied for commuting pairs of elements in $\tilde{\Omega}$.  From Eq.~(\ref{etaProps}) it follows that if $\eta,\eta'\in U$ then $c\eta+c'\eta' \in U$, for all $c,c' \in \mathbb{Z}_d$. Hence $U$ is indeed a vector space, as claimed. 

Key is the relation
\begin{equation}\label{Expans}
A_\Omega^\gamma = \frac{1}{|\tilde{\Gamma}|}\sum_{\tilde{\gamma} \in \tilde{\Gamma}}A_{\tilde{\Omega}}^{\tilde{\gamma}},
\end{equation}
which we now prove, armed with the previous observation. Using the definition of the phase point operators, we start expanding the r.h.s. of Eq.~(\ref{Expans}).
$$
\begin{array}{rcl}
\displaystyle{\frac{1}{|\tilde{\Gamma}|}\sum_{\tilde{\gamma} \in \tilde{\Gamma}}A_{\tilde{\Omega}}^{\tilde{\gamma}}} &=& \displaystyle{\frac{1}{|\tilde{\Gamma}|}\sum_{\tilde{\gamma} \in \tilde{\Gamma}} \frac{1}{d^n}\sum_{a\in \tilde{\Omega}} \omega^{\tilde{\gamma}(a)} T_a}\\
&=& \displaystyle{\frac{1}{|\tilde{\Gamma}|} \frac{1}{d^n}   \sum_{a\in \tilde{\Omega}} \omega^{\tilde{\gamma}_0(a)} T_a \sum_{\eta \in U}  \omega^{\eta(a)}.}
\end{array}
$$
Now we consider two cases. (i) $a\in \Omega$. Then, with property Eq.~(\ref{EP2}),
\begin{equation}\label{SU1}
\sum_{\eta \in U}  \omega^{\eta(a)} = |U|,\;\; \forall a \in \Omega.
\end{equation}
Furthermore, note $|\tilde{\Gamma}|=|U|$.

(ii) $a\in \tilde{\Omega} \backslash \Omega$. There is at least one $\eta\in U$ with $\eta(a) \neq 0$. Since $U$ is a vector space, it follows by character orthogonality that
\begin{equation}\label{SU2}
\sum_{\eta \in U}  \omega^{\eta(a)} = 0,\;\; \forall a \in \tilde{\Omega}\backslash \Omega.
\end{equation}
Inserting Eqs.~(\ref{SU1}) and (\ref{SU2}) in the above expansion, and furthermore using property Eq.~(\ref{EP2}), we find
$$
\frac{1}{|\tilde{\Gamma}|}\sum_{\tilde{\gamma} \in \tilde{\Gamma}}A_{\tilde{\Omega}}^{\tilde{\gamma}} = \frac{1}{d^n} \sum_{a\in \Omega} \omega^{\gamma(a)} T_a = A_\Omega^\gamma.
$$
This proves Eq.~(\ref{Expans}). Now, wlog. we may choose $\tilde{\Omega}$ to be maximal. Since by definition any set $\Omega$ is contained in some maximal set $\tilde{\Omega}(\Omega)$, we may convert any positive state expansion over ${\cal{V}}$ into a positive state expansion over ${\cal{V}}_M$,
$$
\rho = \sum_{\Omega,\gamma} c(\Omega,\gamma) A_\Omega^\gamma = \sum_{\Omega,\gamma}\frac{c(\Omega,\gamma)}{|\tilde{\Gamma}_\Omega|}\sum_{\tilde{\gamma} \in \tilde{\Gamma}_\Omega} A_{\tilde{\Omega}(\Omega)}^{\tilde{\gamma}}
$$
If the expansion coefficients on the l.h.s. are positive, so they are on the r.h.s. $\Box$

\section{Proof of Lemma~\ref{Perpet}}\label{PerpetProof}

{\em{Proof of Lemma~\ref{Perpet}.}} {\em{Statement (A):}} $(\Omega, \gamma + [a,\cdot]) \in {\cal{V}}$, $\forall a\in \Omega$. The set $\Omega$ does not change, and we only need to check the properties in Def.~\ref{PhaSpa} that concern the function update, i.e., Eqs.~(\ref{dgb}), (\ref{NormCond}).

Assume that $\gamma: \Omega \longrightarrow \mathbb{Z}_2$ satisfies $d\gamma = \beta$ on $\Omega$, i.e. $d\gamma(f)=\beta(f)$ for all faces $f\in F(\Omega)$. Consider any such face, with its boundary $\partial f$ consisting of the edges  $c$, $d$ and $c+d$. By definition of $F(\Omega)$ it holds that $c,d,c+d\in \Omega$. Then, with all addition mod 2,
$$
\begin{array}{rcl}
d (\gamma + [a,\cdot])(f) &=& d\gamma(f) + [a,\cdot](\partial f)\\
&=& d\gamma(f) + [a,c] + [a,d] + [a,c+d] \\
&=& d\gamma(f)\\
&=& \beta(f).
\end{array}
$$
Thus, $\gamma+[a,\cdot]$ satisfies Eq.~(\ref{dgb}). 

Furthermore, assume that $\gamma$ satisfies Eq.~(\ref{NormCond}). Then, $\left(\gamma+[a,\cdot]\right)(0)=\gamma(0)+[a,0]=\gamma(0)$. Hence, $\gamma+[a,\cdot]$ satisfies Eq.~(\ref{NormCond}).\smallskip

{\em{Statement (B):}} $(\Omega\times a, \gamma\times s_a) \in {\cal{V}}$, $\forall a\not\in \Omega$ and $s_a \in \mathbb{Z}_2$. There are four items to check in Def.~\ref{PhaSpa}, namely (I) $\Omega \times a$ is closed under inference, (II) $\Omega\times a$ is non-contextual, (III) $\gamma \times s_a$ satisfies Eq.~(\ref{dgb}), and (IV) $\gamma \times s_a$ satisfies Eq.~(\ref{NormCond}).

(I): Consider $c,d \in \Omega \times a$, with $[c,d]=0$, and denote $c'=c+a$, $d'=d+a$. There are three sub-cases.
{\em{(i) $c,d\in \Omega_a$.}} Then, $c+d \in \Omega_a$, since $\Omega_a$ is closed under inference by Lemma~\ref{L1}. Thus, $c+d \in \Omega \times a$.

{\em{(ii) $c\in \Omega_a$, $d\not\in \Omega_a$.}} By construction of $\Omega\times a$, $d' \in \Omega_a$. Thus, $c+d = c+ (d'+a) = (c+d') +a$. Now, since $[c,d]=0$ by assumption and $[c,a]=0$ ($c\in \Omega_a)$ it follows that $[c,d']=0$. Since $\Omega_a$ is closed by Lemma~\ref{L1}, it holds that $c+d' \in \Omega_a$. By construction of $\Omega\times a$, $c+d=(c+d')+a \in \Omega\times a$.

{\em{(iii) $c,d \not \in \Omega_a$.}} By construction of $\Omega\times a$, $c',d'\in \Omega_a$. Thus, $c+d=(c'+a)+(d'+a)=c'+d'$, and further $[c',d']=0$. Since $\Omega_a$ is closed under inference by Lemma~\ref{L1}, $c'+d' =c+d \in \Omega_a$. Thus, $c+d \in \Omega\times a$.

Thus in all three cases, $c,d\in \Omega\times a$, with $[c,d]=0$, implies $c+d \in \Omega \times a$. Hence, $\Omega\times a$ is closed under inference. \smallskip

(III): Assume that $d\gamma =\beta$ on $\Omega$, and consider a triple of edges $c,d,c+d \in \Omega \times a$ with $[c,d]=0$. Then, either (i) all or (ii) one of these edges are in the component $\Omega_a$.

{\em{(i) $c,d,c+d \in \Omega_a$.}} Since $\Omega_a \subset \Omega$ and with Eq.~(\ref{gammaA}), it holds that $d(\gamma\times s_a)(c,d)=d\gamma(c,d)=\beta(c,d)$. 

{\em{(ii)}} W.l.o.g. assume that $c\in \Omega_a$ and $d,c+d \not \in \Omega_a$, and denote $c'=c+a$, $d'=d+a$ as before. Then, for the face $f=(c,d)$ with boundary $\partial f$ consisting of the edges $c$, $d$ and $c+d$, 
$$
\begin{array}{rcl}
d(\gamma\times s_a)(f) &=& \gamma\times s_a(c)+\gamma\times s_a(d)+\\
&&\gamma\times s_a(c+d)\vspace{1mm}\\
&=& \gamma(c) + \gamma(d')+\gamma(c+d')+\\
&& \beta(a,d)+\beta(a,c+d)\vspace{1mm}\\
&=&  \beta(c,d')+\beta(a,d)+\beta(a,c+d)\\
&=& \beta(c,d).
\end{array}
$$
Therein, in the second line we have used Eq.~(\ref{gammaUp}), in the third line Eq.~(\ref{dgb}), in the fourth line Eq.~(\ref{beta2}), and in the fourth line $d\beta(a,d,c)=0$, cf. Eqs.~(\ref{beta_constr}) and (\ref{Def_deb}).\smallskip

(II): Per Def.~(\ref{Def_NC}),  $\Omega \times a$ is non-contextual if there is a function $\tau: \Omega \times a \longrightarrow \mathbb{Z}_2$ that satisfies $d\tau = \beta$. We have explicitly constructed such a function in (III) above, $\tau:= \gamma\times s_a$. 

(IV): Assume that $\gamma:\Omega \longrightarrow \mathbb{Z}_2$ satisfies Eq.~(\ref{NormCond}). Since $0 \in \Omega_a$ for all cnc sets $\Omega$, with Eq.~(\ref{gammaA}) it follows that $\gamma\times s_a(0) = \gamma(0)$, and hence $\gamma\times s_a$ also satisfies Eq.~(\ref{NormCond}). $\Box$

\section{Proof of Lemma~\ref{relR}}\label{ProofRelR}

{\em{Proof of Lemma~\ref{relR}.}} Recall from Lemma~\ref{L4} that each set $\Omega$ can be written in the form $\Omega = \bigcup_{k=1}^{\xi(\Omega)} I_k$, where each $I_k$ is an isotropic subspace, $I_k=\langle a_k,\tilde{I}\rangle$, $a_k\in E$. Therefore, for all $(\Omega,\gamma) \in {\cal{V}}$, it holds that
$$
A_\Omega^\gamma =  \left(\sum_{k=1}^{\xi(\Omega)}A_{I_k}^{\gamma |_{I_k}}\right) - (\xi(\Omega)-1)A_{\tilde{I}}^{\gamma|_{\tilde{I}}}.
$$
Therein, the phase point operators appearing on the r.h.s. are all of the type $m=0$, i.e., they correspond to stabilizer states. 
The Wigner function $\delta_{(\Omega,\gamma)}$ representing the operator $A_\Omega^\gamma$ can thus be expanded as
$$
\delta_{(\Omega,\gamma)} = \left( \sum_{k=1}^{\xi(\Omega)}\delta_{(I_k,\gamma |_{I_k})}\right) - (\xi(\Omega)-1) \delta_{(\tilde{I},\gamma|_{\tilde{I}})}.
$$
Denote by $\|\cdot\|_1$ the 1-norm of the expansion in terms of phase point operators $A_\Omega^\gamma$, and by  $\|\cdot\|_{1,S}$ the 1-norm of the expansion in terms of (density matrices of)  stabilizer states. With the last equation, the triangle inequality, $\| \delta_{(I_k,\gamma|_{I_k})} \|_{1,S} = \| \delta_{(\tilde{I},\gamma|_{\tilde{I}})} \|_{1,S} = 1$, and $\xi(\Omega) \leq 2n+1$ for all cnc sets $\Omega$ (cf. Lemma~\ref{L4}), it follows that
\begin{equation}\label{DelNorm}
\| \delta_{(\Omega,\gamma)} \|_{1,S} \leq 4n+1.
\end{equation}
Now, for any given state $\rho$ consider the optimal representation $W_\rho$, i.e., the one with minimal norm $\|W_\rho\|_1$. Then,
$$
\begin{array}{rcl}
\mathfrak{R}_S(\rho) & \leq & \displaystyle{\|W_\rho\|_{1,S}} = \displaystyle{\left\| \sum_{(\Omega,\gamma) \in {\cal{V}}} W_\rho(\Omega,\gamma) \delta_{(\Omega,\gamma)} \right\|_{1,S}}\vspace{1mm}\\
&\leq& \displaystyle{(4n+1) \sum_{(\Omega,\gamma) \in {\cal{V}}} |W_\rho(\Omega,\gamma)|}\vspace{1mm}\\
& = & \displaystyle{(4n+1) \|W_\rho\|_1}\vspace{1mm}\\
& = & \displaystyle{(4n+1) \mathfrak{R}(\rho).}
\end{array}
$$
Therein, in the first line we have an inequality because the representation $W_\rho$ of $\rho$ is optimized for $\|W_\rho\|_1$, not necessarily for $\|W_\rho\|_{1,S}$. The third line follows by the triangle inequality and Eq.~(\ref{DelNorm}), and the fifth line holds as an equality because $W_\rho$, per assumption, was chosen to minimize $\|W_\rho\|_1$.

This proves the right half of Eq.~(\ref{RobRel}). The left half, $\mathfrak{R}(\rho) \leq \mathfrak{R}_S(\rho)$, follows from the fact that all stabilizer states correspond to phase point operators of type $m=0$. Hence, an expansion in terms of stabilizer states induces an expansion in terms of phase point operators $A_\Omega^\gamma$, with the same non-zero coefficients. $\Box$ 

\section{Computing $W$-representations of many copies of magic states}\label{LTpr} 

Here we describe how to construct valid quasi\-proba\-bilities $W_{\mu^{\otimes n}}$ for $n$ copies of a magic state $\mu$, at bounded computational cost. As with robustness of magic \cite{RoM}, we merge expansions for small numbers of magic states into valid expansions for larger numbers of copies. 

Denote by $\Omega_n^m$ cnc sets $\Omega$ with parameters $n$, $m\leq n$, and choose the phases $\phi$ in Eq.~(\ref{Pauli}) such that
\begin{equation}\label{Tprod}
T_{a+b} = T_a\otimes T_b,\;\;\forall a\in \Omega_{n_1}^{m_1},\, b\in  \Omega_{n_2}^{m_2}.
\end{equation}
Here we identified $a$ and $b$ as elements of $\mathbb{Z}_d^{2(n_1+n_2)}$ by writing $((a_X,0),(a_Z,0))$ and $((0,b_X),(0,b_Z))$, respectively. We then have the following result.
\begin{Lemma}\label{LT}
Be $\Omega_{n_1}^{m_1}$ and $\Omega_{n_2}^0$ two cnc sets with parameters $n_1$, $m_1\leq n$, and $n_2$, $m_2=0$, respectively. Then,
$A_{\Omega_{n_1}^{m_1} \oplus \Omega_{n_2}^0}^\gamma:=A_{\Omega_{n_1}^{m_1}}^{\gamma_1}\otimes A_{\Omega_{n_2}^0}^{\gamma_1}$,
with the function $\gamma: \Omega_{n_1}^{m_1} \oplus \Omega_{n_2}^0 \longrightarrow \mathbb{Z}_2$ defined by
\begin{equation}\label{Decomp}
\gamma(a_1+a_2) := \gamma_1(a_1)+\gamma_2(a_2), 
\end{equation}
for all $a_1 \in \Omega_{n_1}^{m_1}$, $a_2 \in \Omega_{n_1}^0$ is a valid phase point operator on $n_1+n_2$ qubits.
\end{Lemma}

{\em{Proof of Lemma~\ref{LT}.}} We need to verify the properties of Def.~\ref{PhaSpa}, namely (a) that $\Omega_{n_1}^{m_1} \oplus \Omega_{n_2}^0$ is cnc, and (b) that the function $\gamma$ defined in Lemma~\ref{LT} satisfies Eq.~(\ref{dgb}), i.e., $d\gamma = \beta$.

Regarding (a), with Eq.~(\ref{manyI}), $\Omega_{n_1}^{m_1} = \bigcup_{k=1}^{m_1} \langle a_k,\tilde{I}\rangle$, and $\Omega_{n_2}^{0}=I$, with $\tilde{I}$, $I$ isotropic subspaces. Then,
$$
\Omega_{n_1}^{m_1} \oplus \Omega_{n_2}^0 = \bigcup_{k=1}^{m_1}\langle a_k, \tilde{I}\oplus I\rangle,
$$
and $\tilde{I}\oplus I$ is also an isotropic subspace. Hence the set $\Omega_{n_1}^{m_1} \oplus \Omega_{n_2}^0$ is cnc by Lemma~\ref{L4}.\smallskip

Regarding (b), we need to show that for all $a,b\in \Omega_{n_1}^{m_1} \oplus \Omega_{n_2}^0$ with $[a,b]=0$ it holds that $\gamma(a+b)+\gamma(a)+\gamma(b) =\beta(a,b)$. To this end, for any given commuting pair $a$, $b$, we split
$$
a = a_1+a_2,\; b= b_1+b_2,
$$
with  $a_1,b_1 \in \Omega_{n_1}^{m_1}$, $a_2,b_2 \in \Omega_{n_2}^0$. The decompositions are unique. Since $a$ and $b$ commute, $[a_1,b_1]+[a_2,b_2]=0$. Furthermore, $a_2$ commutes with $b_2$, since $\Omega_{n_2}^0$  is an isotropic subspace. Thus,
\begin{equation}\label{ComR}
[a_1,b_1]=[a_2,b_2]=0.
\end{equation}
Further, Eq.~(\ref{Tprod}) implies that
\begin{equation}\label{betaVan}
\beta(c,d) =0,\;\forall c\in  \Omega_{n_1}^{m_1},\; d \in \Omega_{n_2}^0.
\end{equation}
Now we rewrite 
$$
\begin{array}{rcl}
\gamma(a+b) &=& \gamma((a_1+b_1)+(a_2+b_2))\vspace{1mm}\\
&=& \gamma_1(a_1+b_1)+ \gamma_2(a_2+b_2)\vspace{1mm}\\
&=& \gamma_1(a_1)+\gamma_1(b_1) + \gamma_2(a_2) + \gamma_2(b_2) +\\
&&  \beta(a_1,b_1)+\beta(a_2,b_2)\vspace{1mm}\\
&=& \gamma(a_1+a_2)+\gamma(b_1+b_2)+  \beta(a_1,b_1)+\\
&& \beta(a_2,b_2)+\beta(a_1,a_2)+\beta(b_1,b_2) \vspace{1mm}\\
&=& \gamma(a)+\gamma(b)+ \beta(a_1+a_2,b_1+b_2) +\\
&& d\beta(a_1,a_2,b_1+b_2)+d\beta(a_2,b_2,b_1)+\\
&& d\beta(a_1,b_1,a_2+b_2) \vspace{1mm}\\
&=& \gamma(a)+\gamma(b)+ \beta(a,b).
\end{array}
$$
Therein, in the second line we have used the definition of $\gamma$ in Lemma~\ref{LT}. In the third line Eq.~(\ref{ComR}), and in the fourth line the definition of $\gamma$ again. In the fifth line we have used Eq.~(\ref{Def_deb}), Eq.~(\ref{betaSim}), and Eq.~(\ref{betaVan}) on $\beta(a_1+b_1,a_2+b_2)$. In the last line we used Eq.~(\ref{beta_constr}) ($d\beta=0$). $\Box$
\medskip

Denote by $W^{(0)}_\sigma$ an expansion Eq.~(\ref{Wigner}) of the state $\sigma$, but only containing phase point operators with parameter $m=0$, i.e., an expansion into stabilizer states. $W_\rho$ is a valid expansion of $\rho$, according to Eq.~(\ref{Wigner}). Then, it follows from Lemma~\ref{LT} that
\begin{equation}\label{TensDec}
W_{\rho \otimes \sigma}:= W_\rho \otimes W^{(0)}_\sigma
\end{equation}
is a valid expansion of $\rho\otimes \sigma$.

Let $k$ be the largest integer for which decompositions $W_{\mu^{\otimes k}}$ and $W_{\mu^{\otimes k}}^{(0)}$ are obtainable. Then, with Eq.~(\ref{TensDec}),
$W_{\mu^{\otimes n}} = W_{\mu^{\otimes k}} \otimes W_{\mu^{\otimes n-k}} ^{(0)}$, 
and the second factor may be further decomposed as $W_{\mu^{\otimes n-k}} ^{(0)}=W_{\mu^{\otimes k}}^{\otimes (n/k-1)}$, if $k$ divides $n$. Thus, we arrive at an explicit decomposition for $\mu^{\otimes n}$, with 1-norm
$$
\left\|W_{\mu^{\otimes n}}\right\| = \mathfrak{R}\left(\mu^{\otimes k}\right) \mathfrak{R}_S\left(\mu^{\otimes k}\right)^{n/k-1}.
$$
Thus the reduction to blocks of $k$ magic states familiar from the stabilizer case \cite{RoM} can be applied in the present setting as well. By Lemma~\ref{relR}, the resulting 1-norm is lower by a constant factor than of the corresponding expansion into stabilizer states.

\end{document}